\documentclass[12pt,a4paper]{article}
\pdfoutput=1
 \usepackage{jheppub}
 \usepackage{amsmath}
 \usepackage{amsfonts}
 \usepackage{mathtools}
 \usepackage{amssymb}
 \usepackage{caption}
 \usepackage{subcaption}
 \usepackage{slashed}
\usepackage{float}

\def\n{\nu}

\def\k{\kappa}

\def\al{\alpha}

\newcommand{\be} {\begin{equation}}
\newcommand{\ee} {\end{equation}}
\newcommand{\bea} {\begin{eqnarray}}
\newcommand{\eea} {\end{eqnarray}}
\newcommand{\ba} {\begin{array}}
\newcommand{\ea} {\end{array}}
\newcommand{\nn} {\nonumber}

\newcommand{\tew}{\tilde{E_2}}

 \title{${\cal N}=2$ heterotic string compactifications 
 on orbifolds of $K3\times T^2$}
 \author{Aradhita Chattopadhyaya, Justin R. David}
\affiliation{Centre for High Energy Physics, Indian Institute of Science,\\
C. V. Raman Avenue, Bangalore 560012, India.}
\emailAdd{aradhita, justin@cts.iisc.ernet.in}

\abstract{We study ${\cal N}=2$ compactifications of $E_8\times E_8$  heterotic string theory 
on   orbifolds   of $K3 \times T^2$  by $g'$ 
which acts as an  $\mathbb{Z}_N$ automorphism of $K3$ together with a
$1/N$ shift on a circle of $T^2$.  The orbifold action  $g'$  corresponds to the 
$26$  conjugacy classes of the Mathieu group $M_{24}$. 
We show that for the standard embedding  the new supersymmetric index 
for these compactifications  can always be decomposed into  the
 elliptic genus of $K3$ twisted by $g'$. 
 The difference in  one-loop 
corrections  to the gauge couplings are captured by automorphic forms 
obtained by  the theta lifts   of the elliptic genus of $K3$ twisted by $g'$. 
 We work out in detail  the case for which $g'$ belongs to 
the equivalence class $2B$. 
We then investigate all the non-standard embeddings for
$K3$ realized as a $T^4/\mathbb{Z}_\nu$ orbifold with $\nu = 2, 4$ and $g'$ the $2A$ involution. 
We show that for non-standard embeddings
the new supersymmetric index  as well as the difference in one-loop corrections to the 
gauge couplings 
are completely characterized  by the instanton numbers of the embeddings 
together with  the difference in number of hypermultiplets and vector multiplets in the 
spectrum. 
}

\begin{document}
\maketitle
\flushbottom

\section{Introduction}

String compactifications with ${\cal N}=2$ supersymmetry 
has been extensively investigated as  an important  testing ground for string dualities. 
The canonical example of  such a compactification is 
the heterotic string on $K3\times T^2$. 
In the context of string dualities this theory was first investigated in 
\cite{Kachru:1995wm}.  The various theories studied differed on how the spin 
connection was embedded in the gauge connection. 
A simple method of explicitly constructing these compactifications
is to realize $K3$ as a $T^4/ \mathbb{Z}_\nu$ orbifold with  $\nu = 2, 3, 4, 6$. 
A comprehensive list of these orbifold compactifications 
together with all possible embeddings of the spin connection 
in the gauge connection is given in \cite{Stieberger:1998yi,Honecker:2006qz}.
Supersymmetric observables like the new supersymmetric index 
or the  difference in one loop gauge threshold corrections can be shown to be 
independent of the orbifold realization \cite{Harvey-Moore,Cardoso-Curio-Lust,Stieberger:1998yi}.

An important observable in these compactifications 
is the new supersymmetric index 
\cite{Antoniadis:1992sa,Antoniadis:1992rq,Cecotti:1992qh,Cecotti:1992vy,Harvey-Moore,Cardoso-Curio-Lust}
which is defined by 
\begin{equation}
 {\cal Z}_{\rm new} 
  (q, \bar q )  = \frac{1}{\eta^2(\tau)  } {\rm Tr}_{R} 
 \left(    F e^{i \pi  F } q^{L_0 - \frac{c}{24} } \bar q^{\bar L_0 -\frac{ \bar c}{24}}   \right)  \ . 
\end{equation}
Here the  trace  
is performed  over the Ramond sector in  the internal CFT with central 
charges $(c, \bar c) = ( 22, 9 ) $.  
$F$ refers to  the  world sheet fermion number of the right moving 
 ${\cal N}=2$ supersymmetric internal CFT. 
Recently it has been observed that the  new supersymmetric index 
of $K3\times T^2$  which enumerates 
BPS states in these compactifications admits  Mathieu moonshine symmetry \cite{Kachru},
see \cite{Kachru:2016nty} for a review of aspects of moonshine. 
This observation was generalized in \cite{Datta:2015hza}
which considered orbifolds of $K3\times T^2$ 
by $g'$  acted as a $\mathbb{Z}_N$ automorphism in $K3$ and 
and $1/N$ shift on one of the circles of $T^2$. It was observed that  for the standard embedding 
 the new supersymmetric index admits 
a decomposition 
in terms the elliptic genus of $K3$ twisted by $g'$. 
This ensures that the new supersymmetric index admits an expansion 
in terms 
 of the McKay Thompson 
series associated with $g'$ embedded in the Mathieu group $M_{24}$.
It was also observed in \cite{Datta:2015hza} that the difference in 
one loop gauge corrections to gauge couplings
  with Wilson lines for these compactifications 
can be written in terms of Siegel modular forms corresponding to 
the theta lift of the twisted elliptic genus of $K3$. 

The $g'$ considered in these compactifications  of \cite{Datta:2015hza} 
were restricted in the conjugacy class $pA$  of $M_{24}$
with $p=2, 3, 5, 7$. In fact only the class $2A$ was explicitly 
constructed \footnote{We  use the ATLAS naming  for the conjugacy classes
of $M_{24}$ see \cite{Conway}.}, and the analysis was restricted to  
 the standard embedding.
In this paper we  study compactifications  of the $E_8\times E_8$ heterotic 
string theory  on orbifolds
of $K3\times T^2$ by $g'$  in more detail.

We  show that for all $g'$ corresponding to the $26$ conjugacy 
classes of $M_{24}$ and for compactifications  which involve the 
standard embedding of the spin connection of $K3$ into one of the $E_8$'s
the resultant new supersymmetric index always can be written 
in terms of the elliptic genus of $K3$ but twisted by $g'$. 
The standard embedding breaks the gauge group to 
$E_7 \times U(1) \times E_8$. 
The difference in one loop corrections of the gauge groups
$E_7$ and $E_8$ are automorphic forms  of 
$SO(2+s , s; \mathbb{Z} )$  with $s = 0, 1$. 
For $s =0$, the automorphic forms are functions of 
K\"{a}hler, complex structure of the torus $T^2$ while 
for $s=1$ they are also functions of 
the Wilson line embedding in  either of the gauge groups. 
We show that these automorphic forms are obtained as 
theta lifts of the elliptic genus of $K3$ twisted by $g'$. 
We demonstrate these statements explicitly for 2 examples. 
We first consider the situation when $K3$ is realized as $T^4/\mathbb{Z}_4$ and 
then construct the corresponding $g'$ action corresponding to the $2A$ conjugacy class.
We show the new supersymmetric index is determined  by the corresponding 
twisted elliptic genus. This result is identical to that obtained in \cite{Datta:2015hza}
when $K3$ is realized as the orbifold  $T^4/\mathbb{Z}_2$ which illustrates that 
the new supersymmetric index is independent of the realization of $K3$. 
In the second example we consider the situation  when $K3$
is realized as
a rational conformal field theory based on the affine algebra 
$su(2)^6$  and for $g'$ belonging to the conjugacy class
$2B$ studied in \cite{Gaberdiel:2013psa}.  For this situation we show that that
the new supersymmetric index is determined by the elliptic genus of $K3$ twisted 
by the $2B$ action. 

We then examine non-standard embeddings of $K3\times T^2$ compactifications. 
This is done by   considering all the non-standard embeddings
in which $K3$ is realized as a $T^4/Z_2$ as well as 
$T^4/Z_4$  orbifold and the action of $g'$  in  the conjugacy 
class $2A$.  We study the spectrum and then evaluate the 
new supersymmetric index for these compactifications. 
The results  for the spectrum are summarized in tables 
\ref{z2spec}, \ref{spectrumstand2}, \ref{spectrumns1}, \ref{spectrumns2}, \ref{spectrumns3}.
We show   that  the 
new supersymmetric index classifies all the models into 
$4$ distinct types depending on the difference of the number of 
hypermultiplets and vector multiplets, $N_h-N_v$ of the model. 
The result  can be read off using the table \ref{bvalue} and equation (\ref{newindnon})  
  In each case we see that the 
new supersymmetric index again admits  a decomposition 
in terms of the elliptic genus of $K3$ twisted by $g'$.  However there is 
also a dependence in $N_h-N_v$. 
We then evaluate the difference in one loop gauge coupling corrections 
for all these models  with the Wilson line and show that they result in  
$SO(3, 2; \mathbb{Z})$ 
automorphic forms. The automorphic forms for all the  models 
are entirely determined by the instanton numbers of the embeddings 
as well as  $N_h-N_v$ of these models. 
The result can be read off using  the  tables \ref{tblz21}, \ref{tbl1} 
and equation  (\ref{findifthr})

The organization of the paper is as follows. 
In  section \ref{stdembed}  we prove that  for the standard embedding , 
compactifications on orbifolds of $K3\times T^2$ result in 
a new supersymmetric  which  can always  be written in 
terms of the elliptic genus of $K3$ twisted by $g'$.  Section \ref{2examples}
works  out in detail for the situation   when $K3$ is realized as 
$T^4/\mathbb{Z}_4$ with $g' \in 2A$ and 
when  $K3$ is realized as a rational 
conformal field theory based on the $su(2)^6$ affine algebra with $g'\in 2B$. 
In section \ref{nonstdsec} we first introduce all the embeddings 
in which $K3$ is realized 
as a $T^4/\mathbb{Z}_\nu$ orbifold with $\nu =2, 4$ and $g' \in 2A$ 
and evaluate the spectrum, the new supersymmetric index and the 
difference in one loop gauge thresholds. 
 Section \ref{conclus} contains our conclusions. 
 Appendix \ref{notation} contains the notations, conventions and  a list of 
 identities used in the paper, appendix \ref{integrals} contains the 
 details of evaluating one loop threshold integrals. Finally  the appendix \ref{mathfiles}
 summarises the content of mathematica files which were used to arrive
 at some of the results in the paper.

\section{Standard embedding}\label{stdembed}

In this section we first define  ${\cal N}=2$ supersymmetric compactifications of the 
$E_8\times E_8$ heterotic string theory on orbifolds of 
$K3 \times T^2$ by $g'$ in which the spin connection of $K3$ is embedded 
in one of the $E_8$'s in the standard manner. 
$g'$ acts as a $\mathbb{Z}_N$ automorphism of $K3$ together with a
$1/N$ shift along one of the circles of $T^2$. 
The automorphism $g'$ corresponds to  
any of the $26$ conjugacy classes associated with the Mathieu group
$M_{24}$ by which one can twist the elliptic genus of $K3$ 
\cite{Cheng:2010pq,Eguchi:2010fg,Gaberdiel:2010ch}. 

We define the standard embedding   as follows. 
Let the current algebra of one of the $E_8$'s be realized in 
terms of left moving fermions $\lambda^I, I = 1, \cdots 16$. 
The other $E_8$ can be realized in terms of its bosonic lattice or
the fermions $\lambda^{\prime I}$. 
The gauge connection is assumed to have the structure
\begin{equation}
 {\cal G} =  \sum_{I, J  =1}^4 \lambda^{ I}  B_a^{ I  J} \partial X^a  \lambda^  J
 + \sum_{I, J = 5}^{16} \lambda^I A_i^{IJ} \partial X^i \lambda^J 
  + \sum_{I, J=1}^{16}
  \lambda^{\prime I} A_{i} ^{\prime IJ} \partial X^i \lambda^{\prime J} \,.
\end{equation}
Here $A_i, A_i'$ is the flat connection on the $T^2$. 
$B_a$ refers to the $SU(2)$ spin connection of $K3$. 
Thus we have embedded the spin connection in one of the $SU(2)$'s of 
the $E_8$. This  $E_8$ lattice splits into a $D2$ which is coupled to the 
spin connection of $K3$ and a free $D6$ lattice. 
The $D6$ lattice and the second $E_8$ lattice which can contain the 
flat connections $A_i, A_i'$ on  $T^2$  are free. 
Thus we have the $16-4 =12$ free Majorana-Weyl fermions of the 
$D6$ lattice coupled to the flat connection  on the $T^2$ and 
$4$ interacting Majorana-Weyl fermions coupled to the spin 
connection of the $K3$. 
These left moving fermions  with the 
left moving bosons of the $K3$ as well as the right moving supersymmetric 
sector of $K3$ form a $(6, 6)$ conformal field theory. 
Thus the internal CFT of the heterotic string in the standard embedding splits
as 
\begin{equation} \label{decomp}
{\cal H}^{internal} = {\cal H}^{(6, 6)}_{D2K3}
 \otimes  {\cal H}^{(6, 0)}_{D6} \otimes {\cal H} ^{(8 , 0)}_{E_8}  \otimes 
{\cal H}^{(2, 3)}_{T^2} \,.
\end{equation}
Here the second and third Hilbert spaces refer to the $D6$ lattice and the $E_8$ lattice
respectively and the the last refer to the CFT on $T^2$. 
With this decomposition, we can now specify the action of $g'$.  The $g'$ acts as 
a ${\cal Z}_N$ automorphism on the $(6, 6)$ CFT ${\cal H}$ together with a
$1/N$ shift on one of the circles in ${\cal H}^{(2, 3)}_{T^2} $. 

\subsection{New supersymmetric index and  twisted elliptic genus of $K3$} \label{twistind}
Let us now evaluate the new supersymmetric index on the internal CFT
given in (\ref{decomp}). 
\begin{equation} \label{znew}
{\cal Z}_{{\rm new}}=\frac{1}{\eta^2}{\rm Tr}_R((-1)^{F} F q^{L_0-c/24}\bar{q}^{\bar{L_0}-\bar{c}/24}).
\end{equation}
 The right moving Fermion number $F$ can be written as the sum of the 
 Fermion number on $T^2$ together with the Fermion number on $K3$
 \begin{equation}
 F = F^{T^2} +   F^{K3}.
 \end{equation}
 Then it is easy to see that the because of the  right moving Fermion
 zero modes on $T^2$, the only contribution to the index arises 
 from 
\begin{eqnarray}
{\cal Z}_{\rm new} = \frac{1}{\eta^2}{\rm Tr}_R
\left(  F^{T^2} e^{ i \pi (  F^{T^2} +   F^{K3}) } q^{L_0- c/24} \bar q ^{\bar L_0 -\bar{c}/24}  \right) .
\end{eqnarray}
Again examining the trace we can see that the  contributions from left moving 
bosonic and fermionic oscillators on 
$T^2$  cancel. Thus it is only the zero modes on $T^2$ and the left 
moving bosonic oscillators on $T^2$ 
 which contribute 
to the index. 
With these arguments we see that the trace reduces to 
\begin{equation} \label{decompk3}
{\cal Z}_{\rm new} = 
\frac{1}{\eta^2(\tau) }  \frac{ \Gamma_{2, 2}^{(r, s) } ( q, \bar q ) }{\eta^2(\tau)} 
\left [  \frac{\theta_2^6(\tau)}{\eta^6(\tau)}  \Phi^{(r, s) }_R  +
 \frac{  \theta_3^6(\tau)}{\eta^6(\tau) } \Phi^{(r, s) }_{NS^+} 
-   \frac{\theta_4^6(\tau)}{\eta^6(\tau)}  \Phi^{(r, s) }_{NS^-} 
\right]
\frac{E_4(q) }{\eta^8(\tau)}.
\end{equation}
The sum over the sectors $(r, s)$ is implied and  $r, s$  run from $0$ to $N-1$. 
The origin and the definition of each term in the index
is as follows. 
\begin{enumerate}
\item
The  term $\frac{\Gamma_{2, 2}^{(r, s)}}{\eta^2} $ arises from the lattice
sum  on $T^2$ together with the left moving bosonic oscillators. 
The lattice sum is defined as
\begin{eqnarray}
\Gamma_{2, 2}^{(r, s)} (q, \bar q ) &= &
 \sum_{\substack{m_1, m_2, n_2 \in \mathbb{Z}, \\ n_1 = \mathbb{Z} + \frac{r}{N} }}
 q^{ \frac{p_L^2}{2} } \bar q^{\frac{p_R^2}{2}}  e^{2\pi i m_1 s/N} , \\ \nonumber
\frac{1}{2}{p_R^2}  &=& \frac{1}{2  T_2 U_2} | - m_1 U +m_2 + n_1 T + n_2 TU |^2 , \\
  \nonumber
  \frac{1}{2}{p_L^2} &= &\frac{1}{2} p_R^2  + m_1 n_1 + m_2 n_2  \,. 
\end{eqnarray}
$T, U$ are the K\"{a}hler and complex structure of the $T^2$. 
Note that the lattice sum is the only part of the index that contains
anti-holomorphic dependence. Furthermore the insertion of $g'$ and the 
twisted sectors of $g'$ are taken care of by the phase $e^{2\pi i m_1 s/N}$ 
 and the fact the winding modes are shifted from integers by 
 $\frac{r}{N}$. 
\item
The terms in the square bracket arises from evaluating the index on 
the lattice $D6$ together with the combined $D2K3$. 
Note that the partition function on the $D6$ lattice  in the various sectors are given by 
\begin{equation}
 {\cal Z}_R ( D6; q ) = \frac{ \theta_2^6}{\eta^6}, \quad
{\cal Z}_{NS^+} ( D6; q ) = \frac{\theta_3^6}{\eta^6} , \quad
 {\cal Z}_{NS^-} ( D6;  q) = \frac{ \theta_4^6}{\eta^6}. 
 \end{equation}
 While the indices on the combined $D2K3$, $(6, 6)$ conformal field theory 
 are given by 
 \begin{eqnarray} \label{d2k3part}
 \Phi^{(r, s) }_R&=&\frac{1}{N} {\rm Tr}_{R\, R,g^r}[g^s(-1)^{F_R}q^{L_0-c/24}
 \bar{q}^{\bar{L_0} -\bar{c}/24}],\\ \nonumber
 \Phi^{(r, s) }_{NS^+} &=&\frac{1}{N} {\rm Tr}_{NS\, R,g^r}[g^s(-1)^{F_R}q^{L_0 -c/24}
 \bar{q}^{\bar{L_0} -\bar{c}/24 }],\\ \nonumber
\Phi^{(r, s) }_{NS^-} &=&\frac{1}{N} {\rm Tr}_{NS\, R,g^r}[g^s(-1)^{F_R+F_L}
q^{L_0 -c/24 }\bar{q}^{\bar{L_0} -\bar{c}/{24} } ].
 \end{eqnarray}
 We will relate them to the twisted elliptic genus of 
 $K3$ below. 
 \item
 Finally the term $\frac{E_4(q) }{\eta^8(\tau)}$ arises from the partition function of the 
 second $E_8$ which is untouched in the standard embedding. 
 $E_4$ is the Eisenstein series of weight $4$. 
 \end{enumerate}
 
 We now show that the indices in (\ref{d2k3part}) are related to the
 twisted elliptic genus of $K3$ by $g'$. 
 In indices given in (\ref{d2k3part}) 
 note that the spin connection of the $K3$ is coupled to the fermions in $D2$ conformal 
 field theory   and therefore 
 trace can be thought of as a trace in the $K3$ super conformal field theory 
 with central charge $(6, 6)$.  Let us examine 
 the twisted elliptic genus of $K3$ which  is defined as 
 \begin{equation}\label{deftwistep}
 F^{(r,s)}(\tau,z)=\frac{1}{N}{\rm Tr}_{R\, R g'^r}[(-1)^{F_{K3}+\bar{F}_{K3} }g'^s e^{2\pi i zF_{K3}}
 q^{L_0-c/24} \bar{q}^{\bar{L_0}-\bar{c}/24}].
 \end{equation}
 Here $g'$ belongs to to automorphism related to the $26$ conjugacy classes of 
 $M_{24}$.  
 Since this theory admits a  ${\cal N}=2$ spectral flow we can relate the trace over 
 the various sectors in (\ref{d2k3part})  by the following equations
 \begin{eqnarray} \label{relegenus}
 \Phi^{(r, s) }_R &=&  F^{(r,s )}(\tau, \frac{1}{2} ), \\ \nonumber
   \Phi^{(r, s) }_{NS^+} &=&  q^{1/4}  F^{(r,s)}(\tau,  \frac{\tau + 1}{2} ) , \\ \nonumber
   \Phi^{(r, s)}_{NS^-} &=&  q^{1/4}  F^{(r,s)} ( \tau, \frac{\tau}{2} ) .
 \end{eqnarray}
 From (\ref{decompk3}) and (\ref{relegenus}) we see that the 
 new supersymmetric  index for compactifications which involve the 
 standard embedding  admits a decomposition in terms of the 
  elliptic genus of $K3$ twisted by $g'$.  This decomposition then 
 can be used to show that the new supersymmetric index 
 can be expanded in terms of the MacKay-Thompson associated with 
 $g'$  embedded in $M_{24}$ following the arguments of 
 \cite{Kachru,Datta:2015hza}. 

 \subsubsection*{New supersymmetric index in terms Eisenstein series}
 
 Let us further simplify the expression the expression for the 
 new supersymmetric index for the standard embedding. 
 The  elliptic genus of $K3$  twisted by $g'$ in general can be
 written as
 \begin{eqnarray} \label{twistellip}
 F^{(0, 0)} (\tau, z) =   \alpha_{g'}^{(0,0)}   A(\tau, z) , \\ \nonumber
  F^{(0,1)}( \tau, z) = \alpha_{g'}^{(0, 1) }   A (\tau, z)  + \beta_{g'}^{(0, 1) }
  f^{(0, 1)}_{g'} (\tau)
  B  ( \tau, z) ,
 \end{eqnarray}
 where  the Jacobi forms 
 $A( \tau, z) $  and $B( \tau, z)$ are given by 
 \begin{eqnarray}
  A( \tau, z) = \frac{\theta_2 ^2(\tau, z) }{\theta_2^2 (\tau, 0) }  
  + \frac{\theta_3^2(\tau, z) }{\theta_3^2(\tau, 0)}
+ \frac{\theta_4^2 (\tau, z) }{\theta_4^2(\tau, 0)}, 
\qquad 
B(\tau, z) = \frac{\theta_1^2(\tau, z) }{\eta^6 (\tau) }.
  \end{eqnarray}
  The numerical coefficients  $\alpha_{g'}^, \beta_{g'}$ and the 
  form $ f^{(0, 1)}_{g'} (\tau)$ depend on the 
  twist $g'$. For example,  for the conjugacy class $pA$  with $p = 2, 3, 5, 7$ 
  of $M_{24}$ we 
  find 
  \begin{equation}\label{pa}
  \alpha_{pA}^{(0, 0)}  = \frac{8}{p}, \qquad 
   \alpha_{pA}^{(0, 1) }  = \frac{8}{p(p+1)}, \qquad \beta_{pA}^{(0,1)} =  - \frac{2}{p+1}, 
  \end{equation}
  and 
  \begin{equation}
   f^{(0,1)}_{g'}(\tau)   = {\cal E}_p(\tau) =  \frac{12 i }{\pi ( p -1) } \partial_\tau 
   \log \frac{\eta(\tau)}{\eta(p \tau) } \,.
  \end{equation}
  A comprehensive list of the twisted elliptic genus for 
  all the $26$ conjugacy classes of $M_{24}$ can be found in 
  \cite{Eguchi:2010fg}. 
All the remaining elements of the twisted elliptic genus $F^{(r,s)}(\tau, z) $ 
can  be obtained by modular transformations using the relation 
\begin{equation} \label{fulltwist}
 F^{(r, s)} \left( \frac{a\tau + b}{ c\tau + d}, \frac{ z}{ c\tau +d}
 \right) = \exp\left(  2\pi i \frac{c z^2}{ c\tau +d}  \right) 
 F^{( cs + a r, ds + b r)} ( \tau,  z) ,
\end{equation}
with 
\begin{equation}
 a, b, c, d \in \mathbb{Z}, \qquad ad -bc =1.
\end{equation}
In (\ref{fulltwist}) the indices $cs + ar$ and $ds + br $ are taken to be 
mod $N$ where $N$ is the order of $g'$. 
Using this information of the twisted elliptic genus we can write the new supersymmetric 
index for the standard embedding given in (\ref{decompk3}) in terms of Eisenstein series. 
Substituting the following identities 
\begin{align} \label{abid}
 A(\tau, \tfrac{1}{2}  ) &=  \frac{\left( \theta_4^4 \theta_2^2 +  \theta_3^4 \theta_2^2  \right)}{4 \eta^6},     
 \hspace*{2.5cm} B(\tau, \tfrac{1}{2} ) = \frac{\theta_2^2}{ \eta^6},  \\ \nonumber
 A(\tau, \tfrac{\tau +1}{2} ) &= \frac{q^{-1/4} \left( -
    \theta_4^4 \theta_3^2 +  \theta_2^4 \theta_3^2 \right) }{4 \eta^6}, \qquad 
   B(\tau, \tfrac{\tau +1}{2} ) =   \frac{q^{-1/4} \theta_3^2}{ \eta^6}, \\ \nonumber
 A(\tau, \tfrac{\tau}{2}) &=  \frac{q^{-1/4} \left( 
    \theta_3^4 \theta_4^2 +  \theta_2^4 \theta_4^2 \right) }{4 \eta^6} , \qquad 
\ \ \ \ \   B(\tau, \tfrac{\tau}{2} ) = -  \frac{q^{-1/4}  \theta_4^2}{ \eta^6}. 
 \end{align}
in (\ref{decompk3}) and using (\ref{relegenus}) we obtain 
\begin{eqnarray} \label{zneweis}
 {\cal Z}_{\rm new}(q, \bar q )  =-  2 \frac{1}{\eta^{24} } \Gamma_{2,2}^{(r, s) } 
 E_4  \left[ \frac{1}{4}  \alpha^{(r, s) }_{g'}  E_6  -   \beta^{(r, s) }_{g'} f^{(r, s) }_{g'}  E_4
   \right].
\end{eqnarray}
Recall that only the lattice sum is dependent on both $(\tau, \bar \tau) $ while 
the Eisenstein series  $E_6,  E_4$ as well as $f^{(r, s)}$ are holomorphic in 
$\tau$.  Furthermore in the (\ref{zneweis})  sum over $r, s$ from $0, \cdots N-1$
is understood. 

\subsection{Difference of one loop gauge thresholds} \label{thrcorr}

Now let us evaluate the gauge threshold corrections with Wilson line turned on 
in the untouched $E_8$ lattice, we call this gauge group $G$ and the 
broken $E_8$,  $G'$. From the discussion in 
\cite{Cardoso-Curio-Lust}, \cite{Stieberger:1998yi} and \cite{Datta:2015hza}, we see that the new 
supersymmetric index with Wilson line becomes 
\begin{eqnarray} 
{\cal Z}_{\rm new}(q, \bar q ) = - 2  \frac{1}{\eta^{24} } \Gamma_{3,2}^{(r, s) } 
\otimes  E_{4,1}   \left[ \frac{1}{4}  \alpha^{(r, s) }_{g'}  E_6  -   \beta^{(r, s) }_{g'} f^{(r, s) }_{g'}  E_4 \right].
\end{eqnarray}
The presence of the Wilson line introduces an additional moduli $V$ and with $T, U$. 
The lattices sums now are given by 
\begin{eqnarray}\label{defomega}
 \Gamma_{3, 2}^{(r, s)} (q, \bar q ) &= &
 \sum_{\substack{m_1, m_2, n_2, b  \in \mathbb{Z}, \\ n_1 = \mathbb{Z} + \frac{r}{N} }}
 q^{ \frac{p_L^2}{2} } \bar q^{\frac{p_R^2}{2}}  e^{2\pi i m_1 s/N} , \\ \nonumber
 \frac{p_R^2}{2} &=&  \frac{1}{4\, {\rm{det}}{\rm{Im}} \Omega }
 \left| -m_1 U + m_2  + n_1 T + n_2 ( TU - V^2) + b V \right|^2,  \\ \nonumber
 \frac{p_L^2}{2} &=&  \frac{p_R^2}{2} + m_1 n_1  + m_2 n_2  + \frac{1}{4} b^2 , \\ \nonumber
 \Omega &=& \left(  \begin{array}{cc}
                  U & V\\
                  V & T
                 \end{array}
\right) . 
\end{eqnarray}
The product $\otimes$  and function $E_{4,1}$ are defined in the appendix \ref{notation}. 
The one loop corrections to the gauge coupling $G$ is defined by the following integral 
over the fundamental domain
\begin{equation}\label{gthresh}
 \Delta( T, U, V)  = \int_{\cal F} \frac{d^2\tau}{\tau_2} 
 ( {\cal B}_G - b(G) ), 
 \end{equation}
where ${\cal B}$  can be written in terms of the 
new supersymmetric index with the Wilson line as follows
\begin{equation}\label{gthresh1}
 {\cal B}_G = -\frac{2}{24 \eta^{24}} \Gamma^{(r, s)}_{3, 2}
 \otimes \left\{  \tew E_{4, 1} - E_{6, 1} \right\}
  \left[ \frac{1}{4}  \alpha^{(r, s) }_{g'}  E_6  -   \beta^{(r, s) }_{g'}  f^{(r, s) }_{g'}  E_4 \right],
\end{equation}
where 
\begin{equation}
 \tew=  \left( E_2 - \frac{3}{\pi \tau_2} \right).
\end{equation}
The constant $b(G)$ in (\ref{gthresh}) can be fixed by demanding that the integral is well 
defined in the limit $\tau_2\rightarrow \infty$.  The details which are involved in arriving 
at the integrand (\ref{gthresh1}) are given in \cite{Datta:2015hza}  
where the class $2A$ was discussed in 
detail.  Essentially the action of ${\cal B}_G$ is to convert the lattice sum 
with the Wilson line $E_{4, 1}\rightarrow \tew E_{4, 1} - E_{6, 1}$.
This occurs because of is summing over the lattice weighted with the charge vectors. 
Similarly the one loop corrections to the gauge coupling $G'$ is defined 
by an integral of the same form in (\ref{gthresh}), with the integrand given by 
\begin{equation}
 {\cal B}_{G'} = - \frac{2}{24 \eta^{24}}  \Gamma^{(r, s)}_{3, 2}\otimes E_{4, 1} 
 \left[ \frac{1}{4}  \alpha^{(r, s) }_{g'}  \left\{ \tew
 E_6   - E_4^2 \right\} -   \beta^{(r, s) }_{g'}  f^{(r, s) }_{g'}  \left\{ 
 \hat E_2  E_4  -  E_6 \right\} \right].
\end{equation}
Here note that $E_6 \rightarrow \tew E_6 - E_4^2$. 
Using the identities
\begin{eqnarray}\label{ABrel}
\frac{1}{\eta^{24}} ( E_{4,1} ( \tau, z) E_6 - E_{6, 1} ( \tau, z) E_4 ) &=& - 144 B(\tau, z) , 
\\ \nonumber
\frac{1}{\eta^{24}} ( E_{4,1} ( \tau, z) E_4^2 - E_{6, 1} ( \tau, z) E_6 ) &=&  576 A(\tau, z),
\end{eqnarray}
we evaluate  the difference in the one loop thresholds integrands which results in 
\begin{equation}
{\cal B}_G - {\cal B}_{G'} = -12  \Gamma^{(r, s)}_{3, 2}\otimes F^{(r, s)} .
\end{equation}
Thus the difference in the one loop corrections to gauge couplings is given by 
\begin{equation}\label{findiff}
\Delta_G( T, U, V) - \Delta_{G'} ( T, U, V) = -12
\int_{\cal F} \frac{d^2\tau}{\tau_2} \Gamma^{(r, s)}_{3, 2}\otimes F^{(r, s)} .
\end{equation}
There is a constant term that we have ignored 
in the integrand which is necessary to make the integral 
well defined in the $\tau_2\rightarrow \infty$ limit. 

From (\ref{findiff}) we conclude that for compactifications on  the 
orbifold $(K3\times T^2)$ by $g'$ 
involving  the standard embedding, 
the difference in the one loop thresholds is 
the  automorphic form  of $SO(3, 2; \mathbb{Z})$ which is obtained by the 
theta lift of the elliptic genus of $K3$ twisted by $g'$.  To obtain the threshold 
correction without the Wilson line one can take the limit $V\rightarrow 0$ in 
(\ref{findiff}). Then the automorphic form $SO(3, 2; \mathbb{Z})$ reduces
to $SO(2, 2;\mathbb{Z}$ modular forms.

\section{Standard embedding: 2 examples} \label{2examples}

In this section we will discuss in detail 2 examples that demonstrate the 
for   standard embeddings,  the new supersymmetric index can be written in terms of the twisted elliptic 
index. The first example deals with
 the $2A$ orbifold of $K3$ in which $K3$ is at  its $T^4/\mathbb{Z}_4$
limit.   The second example deals with the recent construction  of the $2B$ orbifold of 
$K3$ \cite{Gaberdiel:2013psa}.

\subsection{The $2A$ orbifold from $K3$ as $T^4/\mathbb{Z}_4$}

In this section we will construct the orbifold of $K3$ by $g'$ where $g'$ belongs to the class
$2A$. The well studied method of obtaining this orbifold is 
to realize the $K3$ CFT as a $T^4/\mathbb{Z}_2$ orbifold as discussed in \cite{justin1}. 
Here we will consider the $2A$ orbifold when $K3$ is at the orbifold limit $T^4/\mathbb{Z}_4$. 
As far as we are aware the construction is new. 
This will enable us to investigate the spectrum and 
the threshold corrections of all the non-standard embeddings of heterotic string at the orbifold 
$T^4/\mathbb{Z}_4$ discussed in \cite{Stieberger:1998yi} after the $g'$ action. 

We define the orbifold of $K3$ by $g'$ as follows.  Let us first consider 
$T^4\times T^2$ with co-ordinates $x_1, x_2$ parameterizing $T^2$ and 
$y_1, y_2, y_3, y_4$ labelling $T^4$. 
Then  $K3$ is realized by the  $\mathbb{Z}^4$ which is action given by 
\begin{eqnarray}\label{orb1}
 g^s: & &  \; ( x_1, x_2, y_1+i  y_2, y_3,+i y_4) \sim 
 (x_1, x_2, e^{2\pi i\,s/4}( y_1 + i y_2) ,  e^{-2\pi i\,s/4}(y_3 + i y_4) ), 
 \nonumber \\ 
  & & \qquad\qquad s = 0, 1 ,2 ,3 .
\end{eqnarray}
This orbifold limit of $K3$ is well known and discussed in \cite{Eguchi:1988vra}. 
We now consider the  $g'$ orbifold which is a $\mathbb{Z}_2$ action   given by 
\begin{equation}\label{orb2}
 g': \;   ( x_1, x_2, y_1, y_2, y_3, y_4) \sim  (x_1+\pi, x_2, y_1+\pi, y_2+\pi, y_3 + \pi , y_4 + \pi ).
\end{equation}

We will first show that the twisted  elliptic genus  remains the same as 
that when $K3$ is realized as a $T^4/\mathbb{Z}_2$ orbifold. 
This result in fact a test that the orbifold action 
given in (\ref{orb1}) and (\ref{orb2}) in fact $K3$ twisted by the element
$2A$. We will then evaluate the spectrum of heterotic string compactified 
on this orbifold $K3\times T^2$ for the standard embedding. 
Using the orbifold action we will explicitly show that the new supersymmetric index admits 
a decomposition in terms of the twisted elliptic genus. Therefore this  is 
a verification of the result in the previous section that the 
new supersymmetric index for compactifications  on orbifolds of 
$K3$ in any standard embedding  just depends on  the twisted elliptic 
genus of $K3$. 
We then evaluate the difference in  one loop gauge thresholds and show that indeed 
the resulting modular form is the theta lift of the  elliptic genus of $K3$ twisted by 
the element $2A$.

\subsubsection{Twisted elliptic genus }\label{twistz4ep}

The twisted elliptic genus under under the orbifold  (\ref{orb1}) and (\ref{orb2})  is given by the 
index 
\begin{equation}\label{tgenus}
F^{(r, s) }(\tau, z) =
\frac{1}{8}\sum_{a,b = 0 }^3
Tr_{g^a,g'^r}((-1)^{F_L+\bar{F_R}}g^b g'^s e^{2\pi i z F_L }q^{L_0}\bar{q}^{\bar{L}_0}).\nn
\end{equation}
Here the trace is taken over theory of $4$ free bosonic 
coordinates $y_1, y_2, y_3, y_4$ 
and   $4$ free fermions which form their superpartners, $F_L, F_R$ are the 
left and right moving fermion numbers respectively. We have suppressed the shifts $L_0 -1/4$, 
$\bar L_0- 1/4$  in the definition of the index. 
Let us further define  the trace 
\begin{equation}
 {\cal F}( a, r; b,  s )  = \frac{1}{8}
 Tr_{g^a,g'^r}((-1)^{F_L+\bar{F_R}}g^b g'^s e^{2\pi i z F_L }q^{L_0}\bar{q}^{\bar{L}_0}).
\end{equation}
To evaluate each sector of the above twisted elliptic genus we will 
need the fixed point under the elements $g^ag^{\prime r}$ and what elements 
preserve these fixed points. 
This information is summarized in  table \ref{fixedpoint}.

\begin{table}[H] 
\renewcommand{\arraystretch}{0.5}
\begin{center}
\vspace{-0.1cm}
\begin{tabular}{|c|c|c|c|c|c|c|c|c|}
\hline
 & Fixed points & $g'$ & g & $g^2$ &  $g^3$ & $g'g$ & $g'g^2$ &  $g'g^3$ \\
 \hline
 & & & & & & & & \\
$g$  &0, $\frac{(1+i)}{2}$   & $\times$ & \checkmark & \checkmark &\checkmark &  $\times$ &  $\times$  &  $\times$ \\
\hline
 & & & & & & & & \\
$g^2$  &0, $\frac{(1+i)}{2}$   &  $\times$ & \checkmark & \checkmark &\checkmark &  $\times$  &  $\times$  &  $\times$ \\ \cline{2-9}
 & & & & & & & & \\
 &$\frac{1}{2}$, $\frac{i}{2}$   &  $\times$ &   $\times$ & \checkmark &  $\times$ &  \checkmark  &  $\times$ & \checkmark  \\
\hline
 & & & & & & & & \\
$g^3$  &0, $\frac{(1+i)}{2}$   &  $\times$ & \checkmark & \checkmark &\checkmark &  $\times$ &  $\times$  &  $\times$ \\
\hline
 & & & & & & & & \\
$gg'$  &$\frac{1}{2}$, $\frac{i}{2}$   &  $\times$ &  $\times$ &  \checkmark  &  $\times$ & \checkmark &  $\times$ & \checkmark  \\
\hline
 & & & & & & & & \\
$g^2g'$  &$\frac{1+i}{4}$, $\frac{-1-i}{4}$   &  $\times$ & $\times$ &  $\times$ &  $\times$ &  $\times$ & \checkmark  &  $\times$  \\ \cline{2-9}
 & & & & & & & & \\
 & $\frac{1-i}{4}$, $\frac{-1+i}{4}$  &  $\times$ &   $\times$ & $\times$ &  $\times$ &  $\times$  & \checkmark & $\times$  \\
\hline
 & & & & & & & & \\
$g^3g'$  &$\frac{1}{2}$, $\frac{i}{2}$   &  $\times$ &  $\times$ & \checkmark  &  $\times$ & \checkmark  &  $\times$ & \checkmark \\
\hline
\end{tabular}
\end{center}
\vspace{-0.5cm}
\caption{ Each row lists   the  property of 
fixed points along the $y_1, y_2$ direction  under actions of powers of 
$g, g'$. $\times$ indicates that the fixed point moves, while the $\checkmark$ indicates
the fixed point is invariant. 
Positions are in units of $2\pi$ 
An identical table exists for the $y_3, y_4$ direction. }
\label{fixedpoint}
\renewcommand{\arraystretch}{0.5}
\end{table}

Let us discuss the twisted elliptic genus for each of the sectors. 
The sector $(0, 0)$ is easiest to deal with. Since there are no twists in $g'$ or insertions of $g'$ to deal with 
we  see that the trace reduces to 
\begin{equation}\label{twe0}
F^{0,0}(\tau, z) =\frac{1}{2}Z_{K3} (\tau, z) = 4 A(\tau, z) .
\end{equation}
where $Z_{K3}$ is the elliptic genus of $K3$.

Let us now examine the sector $(0, 1)$.  We see from  table \ref{fixedpoint},
that a single insertion of $g'$ does 
not preserve any of the fixed points. 
Thus we have 
\begin{equation}
 {\cal F} ( a, 0; b, 1) =0, \qquad {\rm for}\;  a = 1, 3.
\end{equation}
Therefore we need to look at ${\cal F} (0, 0; b, 1) $ and ${\cal F}(2, 0; b, 1)$. 
Evaluating the trace in the untwisted sector we see the contributions are 
\begin{eqnarray}
 {\cal F} (0, 0; 0, 1) &=& 0, \qquad  \\ \nonumber
 {\cal F} ( 0, 0; 1, 1) & = & \frac{1}{2}
   \frac{ \theta_1 ( z+ \frac{1}{4} , \tau) \theta_1( - z + \frac{1}{4} ) }
   { \theta_1^2 ( \frac{1}{4} , \tau ) }, \\ \nonumber
 {\cal F} ( 0, 0; 2, 1) & = &  2 
   \frac{ \theta_1 ( z+ \frac{1}{2} , \tau)  \theta_1( - z + \frac{1}{2} ) }{
 \theta_1^2 ( \frac{1}{2} , \tau ) }, \\ \nonumber
 {\cal F} ( 0, 0; 3, 1) & = & \frac{1}{2}
   \frac{ \theta_1 ( z+ \frac{3}{4} , \tau)  \theta_1( - z + \frac{3}{4} ) }{
 \theta_1^2 ( \frac{3}{4} , \tau ) }.
\end{eqnarray}
The numerical coefficients in each of the traces occur due to the contribution of the 
Fermionic zero modes. There are $4$ Fermionic  right moving 
zero modes when $g^2$ is inserted 
in the trace while there are $2$  right moving zero modes for the $g$ and $g^3$ insertions. 
Evaluating the contributions to ${\cal F}(2, 0; b, 1)$ we obtain
\begin{eqnarray} \label{tr1}
{\cal F} ( 2, 0 ; 0 , 1) = 0 , \qquad  {\cal F} ( 2, 0; 2, 1) = 0 , \\ \nonumber
{\cal F}( 2, 0 ; 1, 1) = \frac{1}{2} 
\frac{ \theta_1 ( z+ \frac{2 \tau + 1}{4} , \tau) \theta_1( - z + \frac{2\tau + 1}{4} ) }
   { \theta_1^2 ( \frac{2 \tau + 1 }{4} , \tau ) }, \\ \nonumber
  {\cal F}( 2, 0 ;3, 1) = \frac{1}{2} 
\frac{ \theta_1 ( z+ \frac{2 \tau + 3}{4} , \tau) \theta_1( - z + \frac{2\tau + 3}{4} ) }
   { \theta_1^2 ( \frac{2 \tau + 3 }{4} , \tau ) }. \\ \nonumber
  \end{eqnarray}
  The  vanishing of the first set of equations in (\ref{tr1})  is due to the fact that the 
  fixed points in the relevant traces are not invariant under $g'$ or $g^2 g'$ insertions
  as can be seen from the table \ref{fixedpoint}. 
  The numerical factors in the last line equations in  (\ref{tr1})  is due to 
  presence of $4$ fixed points in these twisted sectors. 
  Now summing up the contributions we obtain 
  \begin{eqnarray} \label{twe1}
  F^{(0, 1) } ( \tau, z) &=& {\cal F} ( 0, 0; 1, 1) +{\cal F} ( 0, 0; 2, 1) + {\cal F}( 0, 0 ;3, 1)
  + {\cal F}( 2, 0 ; 1, 1) +  {\cal F}( 2, 0 ;3, 1),   \nonumber\\
  &=&  4\frac{\theta_2^2(z,\tau)}{\theta_2^2(0,\tau)},  \\ \nonumber
  &=& \frac{4}{3} A( \tau, z)  - \frac{2}{3}  {\cal E}_2(\tau) B(\tau, z) .
  \end{eqnarray}
  The equality in the second line of the above equation is due to identities involving 
  the theta functions. 
  Thus we see that the twisted elliptic genus  of the orbifold given in 
  (\ref{orb1}), (\ref{orb2})  belongs to the class $2A$. 
  
  Though the other sectors  of the twisted
  elliptic genus can be obtained by modular transformations, 
  for completeness we provide some of the details. 
  Lets examine contributions to $F^{(1,0)}$. 
  Due to the presence of right moving Fermionic zero modes we obtain 
  ${\cal F} ( 0, 1; 0, 0 ) =0$. Now the following vanish 
  \begin{eqnarray}
  {\cal F}( 0, 1, a, 0 ) =0  , \qquad {\rm for } \; a = 1, 2, 3, 
  \end{eqnarray}
  This is because due to the 
   insertions of powers of $g$  the trace can contribute only  
   if there are zero modes in the winding sector. However since this sector 
   is twisted in $g'$, the winding modes are all half integer modded and 
   cannot vanish. 
   The only non-trivial  contributions arise from the following
   \begin{eqnarray}
   {\cal F} ( a , 1; b, 0 ) &=& \frac{1}{2} 
   \frac{\theta_1( z + \frac{b + a\tau}{4} ) \theta_1( - z + \frac{ b + a \tau}{4} ) }{
   \theta_1^2 ( \frac{b + a\tau}{4}, z ) }, \qquad {\rm for}\; a =1, 3, \; b = 0, 2, 
   \\ \nonumber
   {\cal F} ( 2, 1;0, 0) &=& 2 
   \frac{ \theta_1 ( z+ \frac{ \tau }{2} , \tau) \theta_1( - z + \frac{\tau }{2}, \tau  ) }
   { \theta_1^2 ( \frac{ \tau  }{2} , \tau ) }.
   \end{eqnarray}
   The rest of the  indices vanish due to the fact that the fixed points in those 
   sectors are not invariant with the relevant insertions of $g, g'$ in the trace. 
   Summing up the contributions it can be seen that 
   \begin{eqnarray} \label{twe2}
   F^{(1,0)}&=&4 \frac{\theta_1(z+ \frac{\tau}{2} , \tau )\theta_1(-z+\frac{\tau}{2}, \tau )}
   {\theta_1(\frac{\tau}{2}, \tau)^2}\\ \nn
&=&4\frac{\theta_4(z,\tau)^2}{\theta_4(0,\tau)^2}.
   \end{eqnarray}

Finally due to the same reasons we see that the only contributions to 
$F^{(1, 1) } $ arise from 
\begin{eqnarray}
{\cal F} (  a, 1; b , 1 ) &=& 
\frac{1}{2} 
   \frac{\theta_1( z + \frac{b + a\tau}{4} ) \theta_1( - z + \frac{ b + a \tau}{4} ) }{
   \theta_1^2 ( \frac{b + a\tau}{4}, z ) }, \qquad {\rm for}\; a =1, 3, \; b = 1, 3, 
   \\ \nonumber
{\cal F} ( 2, 1; 2, 1) &=& 2 
\frac{\theta_1( z + \frac{1 + 1\tau}{4} ) \theta_1( - z + \frac{ 1 + 1 \tau}{4} ) }{
   \theta_1^2 ( \frac{1 + 1\tau}{4}, z ) }.
\end{eqnarray}
Again summing up the contributions leads to 
\begin{eqnarray}\label{twe3}
F^{(1, 1)} & =&   4\frac{\theta_3^2(z,\tau)}{\theta_3^2(0,\tau)^2}\, .
\end{eqnarray}
To conclude, from (\ref{twe0}), (\ref{twe1}), (\ref{twe2}) and (\ref{twe3}) 
we see that the twisted elliptic genus is identical to the class $2A$  first evaluated  
in \cite{justin1} using $K3$ in  the $T^4/\mathbb{Z}_2$ orbifold limit.

\subsubsection{Massless spectrum} \label{standspec}

In this section we will derive the massless  spectrum of heterotic string theory 
compactified on the orbifold given in $g$ in (\ref{orb1}) and $g'$  (\ref{orb2}) with standard embedding. 
In orbifold language the standard embedding of is achieved 
by accompanying the  $\mathbb{Z}_4$ action (\ref{orb1}) together with the 
shift 
\begin{equation} \label{shift}
 V=\frac{1}{4}(1,-1,0^6;0^8) ,
\end{equation}
in the $E_8 \times E_8$ lattice. 
The spectrum of the $T^4/\mathbb{Z}_4$ with the standard shift was first studied in 
\cite{Walton:1987bu}.  We will follow the discussion of \cite{Aldazabal:1995yw} which set up the general discussion 
for studying orbifold compactifications of  heterotic string theory which preserve 
${\cal N}=2$ supersymmetry. 
The orbifold action $g'$  $(\ref{orb2})$ does not produce any fixed points and therefore preserves ${\cal N}=2$
supersymmetry. 
Thus the massless spectrum organizes into the $4$ dimensional 
${\cal N}=2$ gravity multiplet coupled to $N_v$ vectors and $N_h$ hypers. 
The massless states of  the theory  in the $g^n$ twisted sector 
is determined by setting left and right  masses  to zero
\begin{eqnarray}\label{mlmr}
m_L ^2&=&N_L+\frac{1}{2}(P+nV)^2+E_n-1  =0, \\
m_R ^2&=&N_R+\frac{1}{2}(r+nv)^2+E_n-\frac{1}{2} =0.
\end{eqnarray}
Here $P$ is the $E_8\times E_8$ lattice vector which is generically of the form
\begin{equation}
P = ( P_{E_8} ; P_{E_8'} ) .
\end{equation}
The  $8$ dimensional lattice vector $P_{E_8}$ can belong to either the vector or the spinor
conjugacy class  which we denote by 
\begin{eqnarray} \label{conjclass}
\lambda_A&=&(n_1,n_2....n_8)\qquad 
\lambda_B= (n_1+\frac{1}{2},n_{2}+\frac{1}{2}, \cdots, n_{8}+\frac{1}{2} ),\\
{\rm with } & & \sum_{i = 1}^8 n_i =  {\rm even\; integer}. 
\end{eqnarray}
$E_n$ is the shift in the zero point energy on the ground state due to the twisting and is 
given by 
\begin {equation}\label{zpt}
 E_n = \frac{1}{4^2} n ( \nu  - n ) ,
\end {equation}
where $\nu=4$ for the $T^4/\mathbb{Z}_4$ orbifold and $ n = 0, 1, 3, 4 $. 
$r$ is a $SO(8)$ weight  vector  with 
\begin{equation}
\sum_{i=1}^4r_i =\;  {\rm odd},
\end{equation}
$v$ is a $4$ dimensional vector given by 
\begin{equation}\label{defv}
 v = \frac{1}{4} ( 0, 0,  1, 1 ) .
\end{equation}
Further conditions on $r, v, P$ so that we obtain massless states $m_L= m_R =0$ 
will be discussed below. 
The degeneracy of the massless states can be  obtained from \cite{Aldazabal:1995yw}
\begin{eqnarray}\label{unmod}
 D(n) &=&\frac{1}{4 }\sum_{m=0}^{3}\chi(n,m)\Delta(n,m), \\ \nonumber
\Delta(n,m)&=&\exp\left\{2\pi i[(r+nv)mv-(P+nV)mV+\frac{1}{2}mn(V^2-v^2)+m\rho]\right\},
\end{eqnarray}
and $\chi(n, m)$ refers to the number of fixed points in the $g^n$ twisted sector which are invariant 
under the action of $g^m$.  $\rho$  is the phase by which the oscillators in the 
$T^4$ are rotated by the $\mathbb{Z}_4$  action. 
In the untwisted sector $n=0$ we have 
\begin{equation} \label{untw}
\chi (0, m)  =1,
\end{equation}
and the  phases in $D(0)$ simply 
implement the projection of the spectrum under the action of $g^m$. 
From table \ref{fixedpoint} we see that 
\begin{eqnarray} \label{chinon}
 &&\chi( 1 ,  m ) = \chi(3, m ) = 4,  \\ \nonumber
 &&\chi( 2, 0) = 16, \qquad \chi ( 2, 1) = 4, \qquad \chi( 2, 2) = 16, \qquad \chi ( 2, 3) = 4.
\end{eqnarray}

Our goal is to obtain the spectrum when there is a further action  by the $\mathbb{Z}_2$ 
group $g'$ given in (\ref{orb2}).  
The first thing to note is that there are no massless states arising from the 
twisted sectors of $g'$. This is because all these states have  half integer Kaluza-Klein 
modes on $T^4$ and therefore they are massive. 
Thus the only change in obtaining the massless spectrum is that the degeneracy 
given in (\ref{unmod}) changes to 
\begin{equation}\label{dgen2}
D(n;g' ) = \frac{1}{4 }\sum_{m=0}^{3 }\frac{1}{2}\left[ \chi(n,m) + \chi^{(g')} ( n, m ) \right]\Delta(n,m),
\end{equation}
where $\chi^{(g')} $ is the number for fixed points in the $g^n $ twisted sector invariant under the 
action of $g^m g'$. 
Essentially we have inserted the projection over $g'$. 
In the untwisted sector 
\begin{equation}\label{untwg}
\chi^{(g')} ( 0, m )  =  \chi(0,m) =1,
\end{equation} 
and again the phases in 
(\ref{dgen2}) just implement the projection of the spectrum under $g^m$. 
For the twisted sector,  from the tabel \ref{fixedpoint} 
we obtain
\begin{eqnarray} \label{modded}
  &&\chi^{(g')}( 1 ,  m ) = \chi^{(g')} (3, m ) = 0, \\ \nonumber
  && \chi( 2, 0)^{(g')} = 0, \qquad \chi ( 2, 1) = 4, \qquad \chi( 2, 2) = 0, \qquad \chi ( 2, 3) = 4.
\end{eqnarray}

We are now ready to obtain the spectrum of the model. 

\noindent
{\bf Untwisted sector}

It is clear from (\ref{untw}), (\ref{untwg}) and (\ref{dgen2})  we see that there is no change in the 
spectrum for the untwisted sector. 
Thus the untwisted sector remains the same as that worked out earlier in \cite{Aldazabal:1995yw}. 
This sector contains the ${\cal N}=2$ gravity multiplet and the 
${\cal N}=2$ vectors. The gauge group breaks from $E_8\times E_8$ to 
$E_7 \times U(1) \times E_8$ \footnote{We are  ignoring the  $2$   vector multiplets  from 
the one cycles of the $T^2$.}.  Thus the Non-Abelian ${\cal N}=2$ vector multiplets are in the 
${\bf {133}} $ of  $E_7$ and the ${\bf{248}}$ of $E_8$. 
In the untwisted sector there are $2$  singlet hypers  under $E_7 \times E_8$ which 
we denote as $({\bf{1}} , {\bf{1}})$ and 
$2$ hypers charged as $({\bf{56}}, {\bf{1}})$. 

\noindent
The twisted sector consists of only hypermultiplets  

\noindent
{\bf Twisted by $g$ and $g^3$}

From  (\ref{chinon}), (\ref{modded}) and (\ref{dgen2}) we see that the 
degeneracies in the  $g^2$ and $g^3$ twisted sector becomes half of the 
theory on the orbifold $(T^4/\mathbb{Z}_2 ) \times T^2$ worked out in 
\cite{Aldazabal:1995yw}. In fact the states in the $g^3$  twisted sector 
form the anti-particles of the states in the $g$ twisted sector.
The hypers for the $g'$ orbifold are 
$2 (\bf{56}, \bf{1}) + 16 (\bf{1}, \bf{1}) $\footnote{We are not keeping 
track of 
the $U(1)$ charges in our discussion.}. 

\noindent
{\bf  Twisted by $g^2$}

It in only in this sector we really need to explicitly  work out the details of the states  and 
using the formula (\ref{dgen2}). 
For massless states  in the twisted sector we have  the conditions
\begin{equation}\label{condr}
 r^2 = 1, \qquad r\cdot v = -\frac{1}{4}.
\end{equation}
Using the equations (\ref{zpt}), (\ref{defv} and (\ref{condr}) we 
see that $p_R$ given in (\ref{mlmr}) indeed vanishes for $N_R=0$. 
Lets examine  the condition $p_L=0$.  

\begin{enumerate}
\item 
For $N_L=0$ in the $g^2$ twisted sector we see $p_L=0$ results in the 
condition 
\begin{equation}\label{mshell}
(P+2V)^2=3/2.
\end{equation}
This condition can only  be satisfied by two ways. 
Firstly we can take the lattice vectors in both the $E_8$'s  in the vector conjugacy class. 
Thus we have 
\begin{equation}\label{ms1}
(n_1+\frac{1}{2})^2+(n_2-\frac{1}{2})^2+\sum_{j=3}^{16}n_j^2=\frac{3}{2},
\end{equation}
which in turn can be satisfied by $n_1 =0, n_2 =1$ or $n_1 =-1, n_2 =0$ with 
 one of the $n_j =\pm 1, j = 3, 4, 5, 6, 7, 8 $. The restriction that 
 these are in the first lattice comes from the condition  in the last line of 
 (\ref{conjclass}).  All together this results in $24$ solutions. 
 Now the second choice of lattice vectors is,  in which we have 
 the spinor conjugacy class in the first $E_8$ and the vector class
 in the second $E_8$. 
 Therefore (\ref{mshell}) reduces to 
 \begin{equation}\label{ms2}
 (n_1+\frac{1}{2}+\frac{1}{2})^2+
 (n_2+\frac{1}{2}-\frac{1}{2} )^2+\sum_{j=3}^{8}(n_j+\frac{1}{2} )^2+\sum_{k=9}^{16}n_k^2=
 \frac{3}{2} .
 \end{equation}
 Here we can have $n_1=-1$, $n_2=0$ and any odd number of the 
 6 $n_j's$ as 0 or -1 which can be achieved by 32 ways ($^6C_1+^6C_3+^6C_5=32$).
 The $24+ 32 = 56$ solutions of  (\ref{ms1}) and (\ref{ms2}) form the 
 $(\bf{56}, \bf{1}) $ dimensional representation of $E_7\times E_8$. 
 Let us now evaluate the degeneracy of these states. 
 They are  solutions to the mass shell condition  and satisfy $P\cdot V = -1/4$, 
 and have $\rho =0$.  Using (\ref{condr}) and 
 the values of $v$ and $V$ from (\ref{defv}) and (\ref{shift}) respectively 
 We find that 
  $\Delta (2, 1) =1$.  Then from (\ref{dgen2}) we see that the 
  degeneracy of these states is $ D(2, g') = 3$, where we need to divide by $2$ to account 
  for the anti-particles. Thus we have 
  $3 (\bf{56}, \bf{1})$ hypers
   \footnote{For the model just on 
  $T^4/\mathbb{Z}_4  \times T^2$ we have $D(2) = 5$ for these states}.
  \item
  Now lets look at the case of $N_L =1/2$, where the oscillators along the 
  $T^4$ are excited. For these states  there is a pair of oscillators each 
  with $\rho =\pm 1/4$ .
  The $m_L=0$ condition reduces to 
  \begin{equation}
  (P+2V)^2=1/2.
  \end{equation}
  This can be satisfied only when both the $E_8$ lattice vectors are chosen in the 
  vector conjugacy class leading to 
  \begin{equation} \label{2sol}
(n_1+\frac{1}{2} )^2+(n_2-\frac{1}{2} )^2+\sum_{j=3}^{16}n_j^2=\frac{1}{2}.
\end{equation}
 This equation admits two  solutions: $n_1=n_2=n_j=0$ and $n_1=-1, \,n_2=1,\, n_j=0$
which have $P\cdot V =0$. Evaluating the phase  $\Delta(2, 1)$ for $\rho = \pm 1/4 $ we 
obtain $ \Delta(2, 1) = \pm 1 $.  The degeneracy 
 from (\ref{dgen2})  for these states is given by $ 2\times (3+ 1)= 8$, here again we are not 
 counting anti-particles. The $2$ factor arises due to the $2$ solutions for (\ref{2sol}) 
Finally since we have two pairs of oscillators with $\rho = \pm 1/4$ the 
total number of states is given by  have  $2\times 8 = 16$ 
These states are singlets with respect to the $E_7\times E_8$, therefore 
\footnote{For the model 
without the $g'$ orbifold the number of such states is $32$. }. 
\end{enumerate}

To summarize the spectrum of the $g'$ orbifold of $T^4/\mathbb{Z}_4$ with the 
standard shift of (\ref{shift}) consists of a ${\cal N}=2$ gravity multiplet
with a gauge multiplet in the $({\bf 133}, {\bf{248}})$  of  $E_7\times E_8$
and a $U(1)$. The hypermultiplet content is summarized in table \ref{standardspec}. 
Evaluating $N_h - N_v = -12$.  For comparison we have also summarized 
the hypermultiplet content of the same model without the $g'$ model in 
table \ref{standardspec1}. 
The vector multiplet content is the same. 
$N_h - N_v = -244$ for this model which is dictated by anomaly cancellation
since this model admits  a lift to a chiral $6d$ theory unlike the $g'$ orbifold. 
This phenomenon of the vector multiplet being invariant but the reduction 
of the number of hypers by the action of $g'$ was also observed in 
\cite{Datta:2015hza}.  In the subsequent section we will verify that 
the $N_h - N_v = -12$ for the $g'$ orbifold by evaluating the new supersymmetric 
index.

\begin{table}[H]
\renewcommand{\arraystretch}{0.5}
\begin{center}
\vspace{-0.5cm}
\begin{tabular}{|c|c|c|c|c|}
\hline
Model & Shift &Sector & Matter & $N_h-N_v$\\ \hline
 & & $g^0$  & $(\bf{56}, \bf{1} ) +2(\bf{1}, \bf{1} )$ & -12\\ \cline{3-4}
 & & & & \\
$(T^4/\mathbb{Z}_4 \times T^2)/g'$ & $E_7\times U(1)\times E_8$& $g + g^3$
& $2( \bf{56} , \bf{1} )+16( \bf{1}, \bf{1} )$ &\\ \cline{3-4}
 & & & & \\
& $\frac{1}{4}(1,-1,0^6;0^8)$ & $g^2$ & $3( \bf{56}, \bf{1} )+16( \bf{1}, \bf{1} )$ & \\ \hline
\end{tabular}
\end{center}
\vspace{-0.2cm}
\caption{ Hypermultiplet content of the g' orbifold of $T^4/\mathbb{Z}_4 \times T^2$
with the standard embedding. }
\label{standardspec}
\renewcommand{\arraystretch}{0.5}
\end{table}

\begin{table}[H]
\renewcommand{\arraystretch}{0.5}
\begin{center}
\vspace{-0.2cm}
\begin{tabular}{|c|c|c|c|c|}
\hline
Model & Shift &Sector & Matter & $N_h-N_v$\\ \hline
 & & $g^0$  & $(\bf{56}, \bf{1} ) +2(\bf{1}, \bf{1} )$ & +244\\ \cline{3-4}
 & & & & \\
$T^4/\mathbb{Z}_4 \times T^2 $ & $E_7\times U(1)\times E_8$& $g + g^3$
& $4( \bf{56} , \bf{1} )+32( \bf{1}, \bf{1} )$ &\\ \cline{3-4}
 & & & & \\
& $\frac{1}{4}(1,-1,0^6;0^8)$ & $g^2$ & $5( \bf{56}, \bf{1} )+32( \bf{1}, \bf{1} )$ & \\ \hline
\end{tabular}
\end{center}
\vspace{-0.2cm}
\caption{ Hypermultiplet content  of $T^4/\mathbb{Z}_4 \times T^2$
with the standard embedding. }
\label{standardspec1}
\renewcommand{\arraystretch}{0.5}
\end{table}

\subsubsection{The new supersymmetric index}

In this section we will  evaluate the new supersymmetric index for the orbifold 
defined by the actions (\ref{orb1}), (\ref{orb2}) with the shift in (\ref{shift})
in $E_8\times E_8$.  We adapt the method developed in \cite{Stieberger:1998yi} to 
incorporate the additional 
$g'$ orbifolding action. 
Evaluating the trace,  the  new supersymmetric index given in (\ref{znew}) 
splits into the following sectors
\begin{eqnarray} \label{z4std}
{\cal Z}_{\rm new}(q, \bar q)  = - \frac{1}{2\eta^{20}( \tau)  }\sum_{a, b =0}^3 \sum_{r, s =0}^1 
e^{- \frac{2\pi i a b}{16}} 
Z_{E_8 }^{(a, b) } (\tau) \times E_4(q) 
 \times \frac{1}{8 } F( a, r,  b,  s; q) \Gamma^{(r, s)}_{2, 2} (q, \bar q). \nonumber \\
\end{eqnarray}
First note that the anti-holomorphic dependence in $q$ occurs only in the lattice sum 
 $\Gamma^{(r, s)}_{2, 2} (q, \bar q)$
Let us define each of the component in  (\ref{z4std}). 
The trace over the $T^4$ directions is given by 
\begin{equation} \label{fabrs}
F(a, r, b, s; q) = {\rm Tr}_{g^a \, g^{\prime s} R} 
\left( g^b g^{\prime s} e^{i \pi F_{R}^{T^4}} q^{L_0} \bar q ^{\bar L_0}  \right).
\end{equation}
Here the left moving CFT consists of $4$ free bosons with $c=4$ and the 
right movers  consists of $4$ free bosons and $4$ free Fermions which is in the Ramond sector. 
The $F_R$ is the fermion number of the right moving states. 
The explicit expressions for this trace using the orbifold action in 
(\ref{orb1}), (\ref{orb2}) is given by 
\begin{eqnarray} \label{fabrs1}
  F ( a, r, b, s; q)  =
  k^{(a,r,b,s)}\eta^2(\tau)q^{\frac{-a^2}{16}}\frac{1}{\theta_1 ^2(\frac{a\tau+b}{4}, \tau )}.
\end{eqnarray}
The coefficients $k^{(a,r,b,s)}$ for the various values of $(r, s)$ are given by 
the following matrices
\begin{eqnarray}
 k^{(a,0,b,0)}=16\left(\begin{matrix} 0 & 1 & 4 & 1\\ 1 & 1 & 1 & 1\\4 & 1 & 4 & 1\\1 & 1 & 1 & 1
\end{matrix} \right), \quad k^{(a,0,b,1)}=
16\left(\begin{matrix} 0 & 1 & 4 & 1\\ 0 & 0 & 0 & 0\\0 & 1 & 0 & 1\\0 & 0 & 0 & 0
\end{matrix} \right),\\ \nn
k^{(a,1,b,0)}=16\left(\begin{matrix} 0 & 0 & 0 & 0\\ 1 & 0 & 1 & 0\\4 & 1 & 0 & 0\\1 & 0 & 1 & 0
\end{matrix} \right), \quad k^{(a,1,b,1)}=
16\left(\begin{matrix} 0 & 0 & 0 & 0\\ 0 & 1 & 0 & 1\\0 & 0 & 4 & 0\\0 & 1 & 0 & 1
\end{matrix} \right).
\end{eqnarray}
Note that rows and columns are labelled by $a$ and $b$ respectively. The coefficients 
for $(r, s) = (0, 0)$ are identical to the situation without the $g'$ orbifolding. 
The remaining  coefficients can be easily obtained by using the  same  arguments discussed 
in section   while evaluating the twisted elliptic genus of this orbifold. 
The Eisenstein series $E_4(q)$  in (\ref{z4std}) 
results from the partition function of the untouched  $E_8$ lattice 
which is not coupled to the spin connection of $K3$. 
The partition function of the first $E_8$ lattice with the shifts are given by 

\begin{eqnarray} \label{e8part}
\nonumber
Z_{E_8}^{(0,1)}&=&\frac{1}{2}\left\{ \theta_3^6\theta
\left[\begin{smallmatrix}0\\1/2\end{smallmatrix}\right]\theta
\left[\begin{smallmatrix}0\\ -1/2\end{smallmatrix}\right]+
\theta_2^6\theta\left[\begin{smallmatrix}1\\1/2\end{smallmatrix}\right]\theta 
\left[\begin{smallmatrix}1\\-1/2\end{smallmatrix}\right]
+\theta_4^6\theta\left[\begin{smallmatrix}0\\3/2\end{smallmatrix}\right] 
\theta\left[\begin{smallmatrix}0\\-1/2\end{smallmatrix}\right] \right\} \\ \nn
&=&Z_{E_8}^{(0,3)},\\ \nn
Z_{E_8}^{(1,0)}&=&\frac{1}{2}(\theta_3^6\theta\left[\begin{smallmatrix}1/2\\0\end{smallmatrix}\right] \theta\left[\begin{smallmatrix}-1/2\\0\end{smallmatrix}\right]+\theta_2^6 \theta\left[\begin{smallmatrix}3/2\\0\end{smallmatrix}\right] \theta\left[\begin{smallmatrix}-1/2\\0\end{smallmatrix}\right]+\theta_4^6 \theta\left[\begin{smallmatrix}1/2\\1\end{smallmatrix}\right] \theta\left[\begin{smallmatrix}-1/2\\1\end{smallmatrix}\right]) \\ \nn
&=&Z_{E_8}^{(3,0)},\\ \nn
Z_{E_8}^{(1,1)}&=&\frac{1}{2}(\theta_3^6\theta\left[\begin{smallmatrix}1/2\\1/2\end{smallmatrix}\right] \theta\left[\begin{smallmatrix}-1/2\\-1/2\end{smallmatrix}\right]+\theta_2^6 \theta\left[\begin{smallmatrix}3/2\\1/2\end{smallmatrix}\right] \theta\left[\begin{smallmatrix}1/2\\-1/2\end{smallmatrix}\right]+\theta_4^6 \theta\left[\begin{smallmatrix}1/2\\3/2\end{smallmatrix}\right] \theta\left[\begin{smallmatrix}-1/2\\1/2\end{smallmatrix}\right]) \\ \nn
&=&-Z_{E_8}^{(3,3)},\\ \nn
Z_{E_8}^{(1,2)}&=&\frac{1}{2}(\theta_3^6\theta\left[\begin{smallmatrix}1/2\\1\end{smallmatrix}\right] \theta\left[\begin{smallmatrix}-1/2\\-1\end{smallmatrix}\right]+\theta_2^6 \theta\left[\begin{smallmatrix}3/2\\1\end{smallmatrix}\right] \theta\left[\begin{smallmatrix}1/2\\-1\end{smallmatrix}\right]+\theta_4^6 \theta\left[\begin{smallmatrix}1/2\\2\end{smallmatrix}\right] \theta\left[\begin{smallmatrix}-1/2\\0\end{smallmatrix}\right]) \\ \nn
&=&-Z_{E_8}^{(3,2)},\\ \nn
Z_{E_8}^{(1,3)}&=&\frac{1}{2}(\theta_3^6\theta\left[\begin{smallmatrix}1/2\\3/2\end{smallmatrix}\right] \theta\left[\begin{smallmatrix}-1/2\\-3/2\end{smallmatrix}\right]+\theta_2^6 \theta\left[\begin{smallmatrix}3/2\\3/2\end{smallmatrix}\right] \theta\left[\begin{smallmatrix}1/2\\-3/2\end{smallmatrix}\right]+\theta_4^6 \theta\left[\begin{smallmatrix}1/2\\5/2\end{smallmatrix}\right] \theta\left[\begin{smallmatrix}-1/2\\-1/2\end{smallmatrix}\right]) \\ \nn
&=&-Z_{E_8}^{(3,1)},\\ \nn
Z_{E_8}^{(2,1)}&=&\frac{1}{2}(\theta_3^6\theta\left[\begin{smallmatrix}1\\1/2\end{smallmatrix}\right] \theta\left[\begin{smallmatrix}-1\\-1/2\end{smallmatrix}\right]+\theta_2^6 \theta\left[\begin{smallmatrix}2\\1/2\end{smallmatrix}\right] \theta\left[\begin{smallmatrix}0\\-1/2\end{smallmatrix}\right]+\theta_4^6 \theta\left[\begin{smallmatrix}1\\3/2\end{smallmatrix}\right] \theta\left[\begin{smallmatrix}-1\\-1/2\end{smallmatrix}\right]) \\ \nn
&=&Z_{E_8}^{(2,3)}.\\ 
\end{eqnarray} 
Also in the $\mathbb{Z}_2$ subgroup sector we have
\begin{eqnarray}
Z_{E_8}^{(0,2)}&=&\frac{1}{2}(\theta_3^6\theta\left[\begin{smallmatrix}0\\1\end{smallmatrix}\right] \theta\left[\begin{smallmatrix}0\\-1\end{smallmatrix}\right]+\theta_4^6\theta\left[\begin{smallmatrix}0\\2\end{smallmatrix}\right] \theta\left[\begin{smallmatrix}0\\0\end{smallmatrix}\right])\\ \nn
&=&\frac{1}{2}(\theta_3^6\theta_4^2+\theta_4^6\theta_3^2) ,\\ \nn
Z_{E_8}^{(2,0)}&=&\frac{1}{2}(\theta_3^6\theta\left[\begin{smallmatrix}1\\0\end{smallmatrix}\right] \theta\left[\begin{smallmatrix}-1\\0\end{smallmatrix}\right]+\theta_2^6\theta\left[\begin{smallmatrix}2\\0\end{smallmatrix}\right] \theta\left[\begin{smallmatrix}0\\0\end{smallmatrix}\right])\\ \nn
&=&\frac{1}{2}(\theta_3^6\theta_2^2+\theta_2^6\theta_3^2) ,\\ \nn
Z_{E_8}^{(2,2)}&=&\frac{1}{2}(\theta_4^6\theta\left[\begin{smallmatrix}1\\2\end{smallmatrix}\right] \theta\left[\begin{smallmatrix}1\\0\end{smallmatrix}\right]+\theta_2^6\theta\left[\begin{smallmatrix}2\\1\end{smallmatrix}\right] \theta\left[\begin{smallmatrix}0\\-1\end{smallmatrix}\right]) \\ \nn
&=&\frac{1}{2}(-\theta_4^6\theta_2^2+\theta_2^6\theta_4^2) .\\ \nn
\end{eqnarray} 
The definition of  the generalized Jacobi theta functions is given by 
\begin{equation}
 \theta\left[\begin{smallmatrix}a\\ b\end{smallmatrix}\right](\tau ,z)
=\sum_{k\in \mathbb{Z}}q^{\pi i \tau (k+\frac{a}{2})^2}e^{\pi i (k+\frac{a}{2})b}e^{ 2\pi i ( k+\frac{a}{2}) }.
\end{equation}
Note that  $\theta_1(\tau ,z)=\theta\left[\begin{smallmatrix}1\\ 1\end{smallmatrix}\right](\tau ,z)$
In the above equation when the argument of the $\theta$-function is not explicitly mentioned, 
it is understood that it  is evaluated at $z=0$ and at $\tau$.

We can now sum over $(a, b)$ in the equation (\ref{z4std}). 
After using (\ref{fabrs}) and (\ref{e8part}) we obtain the expected results
\begin{eqnarray}\label{z4std1}
 & & {\cal Z}_{\rm new} ( q, \bar q ) = - \frac{2}{\eta^{24} (\tau) } 
 \sum_{r, s =0}^1
 \Gamma^{(r, s)}_{2, 2}
 E_4 \left[\frac{1}{4} \alpha_{2A}^{(r, s) } E_6 - \beta_{2A}^{r, s} f_{2A}^{(r, s)} (\tau) E_4 \right],
 \\ \nonumber
 & & \alpha_{2A}^{(0, 0)} =  4 ,\qquad  \beta_{2A}^{(0,0) } =0 , \\ \nonumber
 & & \alpha_{2A}^{ (0, 1)} = \frac{4}{3}, \qquad \beta_{2A}^{(0,1)} = -\frac{2}{3} , \\ \nonumber
 & & \alpha_{2A}^{(1, 0)} =   \alpha_{2A}^{(1, 1)}=  \frac{4}{3} ,\qquad 
 \beta_{2A} ^{(1, 0)} =   \beta_{2A} ^{(1, 1)} = \frac{1}{3}, \\ \nonumber
 & & f_{2A}^{(0, 1)} (\tau) = {\cal E}_2(\tau) , \quad
  f_{2A}^{(1, 0)} (\tau) =    {\cal E}_2( \frac{\tau}{2} ) , \quad 
  f_{2A}^{(1, 1)} (\tau) = {\cal E}_2( \frac{ \tau +1}{2} ). 
\end{eqnarray}
We performed the sum over $(a, b)$ in (\ref{z4std}) for each of the $(r, s)$ sectors using Mathematica
to arrive at the result (\ref{z4std1}). 

From (\ref{pa}) we see that 
the new supersymmetric index  of the orbifold of $T^4/\mathbb{Z}_4 \times T^2$ by $g'$ 
agrees with that of the $2A$ orbifold of $K3\times T^2$. 
This result was expected  since we have seen in section \ref{twistz4ep}, that 
the twisted elliptic genus of the orbifold in (\ref{orb1}), (\ref{orb2}) 
agrees with the $2A$ class. Then  the general arguments in section \ref{twistind}
show that for  standard embeddings the new supersymmetric index can be written in 
terms of the twisted elliptic genus. 
However it is indeed nice to see this using explicit computations.

As a consistency check of our calculations we will evaluate the 
$N_h -N_v$ from the new supersymmetric index. 
From the general arguments of \cite{Harvey-Moore} the    $q^{1/6}$   coefficient of 
the following expression which is related to the new supersymmetric index
evaluates $N_h - N_v$. 
\begin{equation}\label{nhnv}
 N_h - N_v = \left.  \frac{1}{4} \eta^4 
 \left( \sum_{ s=0}^{N} {\cal Z}_{\rm new}^{(0, s)} \right)  \right|_{q^{1/6}} ,
\end{equation}
where ${\cal Z}_{\rm new}^{(0, s)}$ is the corresponding sector of the
new supersymmetric index without the lattice factor $\Gamma^{(0,2)}_{2,2}$. 
We focus on these terms to extract out the massless states contributing 
to the new supersymmetric index.  The $\frac{1}{4}$ factor is 
introduced to take into account the normalizations of the new supersymmetric
index used in this paper. 
Substituting the new supersymmetric index for the standard 
embedding of the $2A$ orbifold of $K3\times T^2$ evaluated in 
(\ref{z4std1}) we obtain 
\begin{equation}
(N_h - N_v)|_{2A}  = - 12 .
\end{equation}
Note that this agrees with the explicit computation of the spectrum in 
table \ref{standardspec} \footnote{We have evaluated $(N_h -N_v)$ from the 
new supersymmetric index for all the $pA$ orbifolds of $K3\times T^2$ with 
$p = 3, 5, 7, 11$. We  obtain  $-134, -256, -317, -376 $ respectively which 
indicates that the number of hypers is reduced by this orbifolding. 
It is also an important check on the 
compactification that  we obtain integers in all these situations. } .

Now turning on Wilson line  in the unbroken $E_8$ and evaluating the thresholds proceeds identically 
to that discussed in section \ref{thrcorr}. We thus obtain the result that the difference in one loop 
gauge thresholds for this orbifold compactification is the theta lift of the 
twisted elliptic genus of $K3$ belonging to the class $2A$.

\subsection{The  $2B$ orbifold from $K3$ based on ${\rm su}(2)^6$. }
\label{su2bm}

Recently in \cite{Gaberdiel:2013psa},  the $K3$ sigma model has been studied 
in terms of a rational conformal field theory based on the affine algebra ${\rm su} (2)^6$. 
In this model of $K3$ the action of $g'$ 
\footnote{In \cite{Gaberdiel:2013psa}, 
$g'$ was referred to as $g$, see section 6.1.} an element of order $4$, 
which belongs to the conjugacy class
$2B$ of $M_{24}$ was explicitly constructed  and 
the twisted elliptic genus was evaluated. 
In this section we will use this realization of $K3$ to evaluate the new supersymmetric 
index of heterotic compactified on $K3\times T^2$ orbifolded by the 
order $4$ element $g'$. 
We will show that  indeed as demonstrated  by the general analysis of section \ref{twistind}, 
that new supersymmetric index can be written in terms of the twisted 
elliptic genus of $K3$ twisted by $g'$. Furthermore  as discussed in 
section \ref{thrcorr}, this implies that
the difference in one loop gauge thresholds is determined by the 
theta lift of the corresponding twisted elliptic genus.

\subsubsection{Twisted elliptic genus} \label{sutwelip}

Let us evaluate the twisted elliptic genus as 
defined by the trace in (\ref{deftwistep}). 
From the definition of the trace we need the characters of the ${\rm su}(2)^6$ model in the Ramond section. 
These were listed in \cite{Gaberdiel:2013psa}, here we present them in the 
table \ref{twistedeg}. 
\begin{table}[H] 
\renewcommand{\arraystretch}{1.0}
\begin{center}
\vspace{0.5cm}
\begin{tabular}{|c|c|c|}
\hline
$R^-$ & [10 00 00, 10 00 00] & -[01 11 11, 01 00 00]\\
&  [01 00 00, 01 00 00] & -[10 11 11, 10 00 00]\\
&  [00 10 00, 00 10 00] & -[11 01 11, 00 10 00]\\
&  [00 01 00, 00 01 00] & -[11 10 11, 00 01 00]\\
&  [00 00 10, 00 00 10] & -[11 11 01, 00 00 10]\\
&  [00 00 01, 00 00 01] & -[11 11 10, 00 00 01]\\
\hline 
\end{tabular}
\end{center}
\vspace{-0.2cm}
\caption{${\rm su}(2)^6$ characters in the Ramond sector with the  sign 
$(-1)^{F_L + F_R} $}  \label{twistedeg}
\renewcommand{\arraystretch}{1.0}
\end{table}
\noindent
${\rm su}(2)_k$ characters  of the highest weight representation $[a]$ with 
$ a=0,...k$ are given by 
\be
{\rm ch}_{k,\frac{a}{2}}(\tau,z)={\rm Tr}_{[a]_k}q^{L_0-c/24} e^{ 2\pi i z J_0} \, .
\ee
Thus $0$   in table \ref{twistedeg} represents the   ${\rm su}(2)$ character at level 1 
\begin{equation}\label{cho}
 {\rm ch}_{1,0}=\frac{\theta_3(2\tau,2z)}{\eta(\tau)},
\end{equation}
while $1$ represent the spinorial ${\rm su}(2)$  character given by
\begin{equation}\label{ch1}
 {\rm ch}_{1,\frac{1}{2}}=\frac{\theta_2(2\tau,2z)}{\eta(\tau)}.
\end{equation}
The comma in the list of  table \ref{twistedeg}  separates the left moving ${\rm su}(2)$   characters 
and the right moving ones. 
The $SU(2)_L \times SU(2)_R$   $R$-symmetry of $K3$ is carried by the first 
${\rm su}(2)$ character among the left and right moving characters respectively.
As shown in \cite{Gaberdiel:2013psa}, the elliptic genus with the characters given in 
the table reduces to that of $K3$. 

The $g'$ orbifold on $K3$ is implemented by the action 
\begin{equation} \label{gprime}
g'= \rho_L 
\left[ 
\left( \begin{matrix} 1 & 0\\ 0 & 1 \end{matrix} \right)
\left( \begin{matrix} -1 & 0\\ 0 & -1 \end{matrix} \right)
\left( \begin{matrix} i & 0\\ 0 & -i \end{matrix} \right)
\left( \begin{matrix} -i & 0\\ 0 & i \end{matrix} \right)
\left( \begin{matrix} -i & 0\\ 0 & i \end{matrix} \right)
\left( \begin{matrix} -i & 0\\ 0 & i \end{matrix} \right) \right] .
\end{equation}
Where $\rho_L$ refers to the fact that the action of $g'$ is restricted to the 
left moving characters. 
The $SU(2)$ rotation  matrices of  $g'$  on  the ${\rm su}(2)$ characters is given by 
\begin{eqnarray}
 \label{gprimesu2}
{\rm Tr}_{[0]} \left[ 
\left( \begin{matrix} -1 & 0\\ 0 & -1 \end{matrix} \right)q^{L_0-\frac{1}{24}}
e^{2\pi  i J_0} \right] &=&\frac{\theta_3(2\tau,2z)}{\eta(\tau) },\\ \nn
{\rm Tr}_{[1]} \left[ 
\left( \begin{matrix} -1 & 0\\ 0 & -1 \end{matrix} \right)q^{L_0-\frac{1}{24}}
e^{ 2\pi i J_0} \right] &=&-\frac{\theta_2(2\tau,2z)}{\eta(\tau) },\\ \nn
{\rm Tr}_{[0]} \left[ 
\left( \begin{matrix} i & 0\\ 0 & -i \end{matrix} \right)q^{L_0-\frac{1}{24}}
e^{2\pi i J_0}\right] &=&\frac{\theta_4(2\tau,2z)}{\eta(\tau) },\\ \nn
{\rm Tr}_{[1]} \left[ 
\left( \begin{matrix} i & 0\\ 0 & -i \end{matrix} \right)q^{L_0-\frac{1}{24}}
e^{2\pi i J_0} \right] &=&-\frac{\theta_1(2\tau,2z)}{\eta(\tau) } .
\end{eqnarray}

The $F^{(0, 0)}$ component of the elliptic genus is easy to evaluate and 
we see that it is given by 
\begin{eqnarray}
 F^{0,0} (\tau, z)&=&\frac{1}{2 \eta^6(\tau) }
 \left[ \theta_2(2\tau,2z)\theta_3(2\tau)^5-\theta_3(2\tau,2z)\theta_2(2\tau)^5 \right. 
 \\ \nn
&&\left. 
+5\theta_3(2\tau,2z)\theta_2(2\tau)\theta_3(2\tau)^4-5\theta_2(2\tau,2z)\theta_3(2\tau)\theta_2(2\tau)^4 \right] 
\\ \nonumber
& =& 2  A(\tau, z) .
\end{eqnarray}
On evaluating the trace, the right movers contribute a factor of $2$ since the zero modes  form a 
$SU(2)$ doublet. 
Note that the $F^{(0,0)}$,  component differs from the elliptic genus of $K3$ by a $1/4$ factor. 
Using  the action of $g'$ on the characters we evaluate  the following  components of the 
twisted elliptic genus to be 
\begin{eqnarray} \nonumber
 F^{(0,1) }(\tau, z) &=&\frac{1}{2 \eta^6 (\tau) }\left[ \theta_2(2\tau,2z)\theta_3(2\tau)\theta_4(2\tau)^4-
\theta_3(2\tau,2z)\theta_2(2\tau)\theta_4(2\tau)^4 \right] \\ \nonumber
&=& \frac{1}{2} \left[  {\cal E}_2 ( \tau) - 2 {\cal E}_4 (\tau)  \right]  B(\tau , z) , \\ \nonumber
F^{(0, 2) } (\tau, z) &=& 
\frac{1}{2 \eta^6(\tau) }
\left[ 
\theta_2(2\tau,2z)\theta_3(2\tau)^5-\theta_3(2\tau,2z)\theta_2(2\tau)^5 \right.  \\ \nn
&& \left. 
-3\theta_3(2\tau,2z)\theta_2(2\tau)\theta_3(2\tau)^4+3\theta_2(2\tau,2z)\theta_3(2\tau)\theta_2(2\tau)^4 \right]
 \\ 
&=& - \frac{2}{3} \left[  A(\tau, z) + {\cal E}_2(\tau) B(\tau, z)  \right].
\end{eqnarray}
All the remaining components of the twisted elliptic genus can be  obtained from modular 
transform given in (\ref{fulltwist}). 
Note that the twisted elliptic genus falls into the form given in (\ref{twistellip}) with the identifications
\begin{eqnarray} \label{alphabeta}
& &  \alpha_{2B}^{(0, 0)} = 2, \qquad \alpha_{2B}^{(0, 1)} = 0, \quad \alpha_{2B}^{(0,2)} = -\frac{2}{3}, 
 \\ \nonumber
 & & \beta_{2B}^{(0, 1)} = \frac{1}{2} , \qquad f_{2B}^{(0, 1)} =  {\cal E}_2(\tau) - 2 {\cal E}_4(\tau) , 
 \\ \nonumber
 & &\beta_{2B}^{(0, 2)} = - \frac{2}{3} , \qquad f_{2B}^{(0, 2)} =  {\cal E}_2(\tau) .
\end{eqnarray}

\subsubsection{ New Supersymmetric Index}

From the discussion in section \ref{sutwelip} in which $K3$ is realized as a rational ${\rm su}(2)^6$ rational 
conformal field theory we see that the $R$ symmetry of the model is carried by the 
first character among both the left and right movers.
The new supersymmetric  index given in (\ref{znew}) involves the trace in which the 
right movers are always in the Ramond sector with a $(-1)^{F_R}$. The right moving characters listed 
in the table \ref{twistedeg} are indeed in the  $R^-$ sector. 
The standard embedding identifies  $R$ symmetry of the left movers carried by the first character of 
in the ${\rm su}(2)^6$ model  with the  fermions of the $D2$ lattice in the first $E_8$.  
Now from the expression of the new supersymmetric index in 
(\ref{decompk3}) we see one needs this first character in the $R^+, NS^+$ and $NS^-$ sectors. 
These sectors couple  to the corresponding sectors of the $D6$ lattice realized in terms of 
fermions.  Table \ref{fznew}  lists the characters the $R^+, NS^+$ and $NS^-$ of the 
${\rm su}(2)^6$ CFT. 
\begin{table}[H] 
\renewcommand{\arraystretch}{1.0}
\begin{center}
\vspace{-0.2cm}
\begin{tabular}{|c|c|c|}
\hline
$R^+$ & -[10 00 00, 10 00 00] & -[01 11 11, 01 00 00]\\
&  [01 00 00, 01 00 00] & [10 11 11, 10 00 00]\\
&  [00 10 00, 00 10 00] & [11 01 11, 00 10 00]\\
&  [00 01 00, 00 01 00] & [11 10 11, 00 01 00]\\
&  [00 00 10, 00 00 10] & [11 11 01, 00 00 10]\\
&  [00 00 01, 00 00 01] & [11 11 10, 00 00 01]\\
\hline
$NS^-$ & [00 00 00, 10 00 00] & -[11 11 11, 01 00 00]\\
&  [11 00 00, 01 00 00] & -[00 11 11, 10 00 00]\\
&  [10 10 00, 00 10 00] & -[01 01 11, 00 10 00]\\
&  [10 01 00, 00 01 00] & -[01 10 11, 00 01 00]\\
&  [10 00 10, 00 00 10] & -[01 11 01, 00 00 10]\\
&  [10 00 01, 00 00 01] & -[01 11 10, 00 00 01]\\
\hline
$NS^+$ & -[00 00 00, 10 00 00] & -[11 11 11, 01 00 00]\\
&  [11 00 00, 01 00 00] & [00 11 11, 10 00 00]\\
&  [10 10 00, 00 10 00] & [01 01 11, 00 10 00]\\
&  [10 01 00, 00 01 00] & [01 10 11, 00 01 00]\\
&  [10 00 10, 00 00 10] & [01 11 01, 00 00 10]\\
&  [10 00 01, 00 00 01] & [01 11 10, 00 00 01]\\
\hline 
\end{tabular}
\end{center}
\vspace{-0.5cm}
\caption{$\widehat{\rm su}(2)^6$ characters in sectors relevant of 
evaluating ${\rm Z}_{new}$. } \label{fznew}
\renewcommand{\arraystretch}{1.0}
\end{table}
Comparing tables (\ref{fznew}) and (\ref{twistedeg}) we can see how the spinor representations
of the first character in the left moving sector has become a scalar character when the 
Ramond sector flows to the Neveu-Schwarz sector. 

Let us first evaluate the component $\Phi^{(0,0)}$ in various sectors. 
Using the character table \ref{fznew} and the  rules in (\ref{cho}) and (\ref{ch1}) we
obtain
\begin{eqnarray}
 \Phi^{(0, 0)}_{R^+} &=&
 \frac{1}{2\eta(\tau)^6}(4\theta_3^5(2\tau)\theta_2(2\tau)+4\theta_2^5(2\tau)\theta_3(2\tau)), \\ \nonumber
 &=& \frac{1}{2}[\frac{\theta_2^2}{\eta^6}(\theta_3^4+\theta_4^4)], \\ \nonumber
 \Phi^{(0, 0)}_{NS^-} &=& 
 \frac{1}{2 \eta(\tau)^{6}}
[5\theta_2^2(2\tau)\theta_3^4(2\tau)-5\theta_3^2(2\tau)\theta_2^4(2\tau)+\theta_3^6(2\tau)-\theta_2^6(2\tau)], 
\\ \nonumber
&=& \frac{1}{2}[\frac{\theta_3^2}{\eta^6}(\theta_2^4-\theta_4^4)], \\ \nonumber
\Phi^{(0,0)}_{NS^+} &=& 
\frac{1}{2\eta(\tau)^{6}}
[5\theta_2^2(2\tau)\theta_3^4(2\tau)+5\theta_3^2(2\tau)\theta_2^4(2\tau)-\theta_3^6(2\tau)+\theta_2(2\tau)^6], 
\\ \nonumber
&=& \frac{1}{2}[\frac{\theta_3^2}{\eta^6}(\theta_2^4-\theta_4^4)].
\end{eqnarray}
Here we have used Riemann's bilinear identities to simplify the resulting expressions 
and obtain the result in terms of theta functions with argument $\tau$. 
We can now multiply these along with the characters of the $D6$ lattice in the 
corresponding sectors as  given in (\ref{decompk3}) and we obtain 
the following result for the $(0, 0)$ sector of the new supersymmetric index
\begin{eqnarray}\label{z00}
 {\cal Z}_{\rm new}|_{(0, 0)} = - 2 \frac{1}{\eta^{24} (\tau) }
 \Gamma^{(0, 0)}_{2, 2} \times \frac{2}{4} E_4 E_6 \, .
\end{eqnarray}
Note that this is  $\frac{1}{4}$ of 
the result expected for compactifications of heterotic on $K3\times T^2$. 
Lets move now to the $(0, 1)$ sector which represents a single insertion of $g'$. 
For $\Phi^{(0,1)}_{R^+}$  using the results in (\ref{gprimesu2})  for the characters with a single insertion
of $g'$ we see that the only characters which survive are  $-[10 00 00, 10 00 00]$ and  $[01 00 00, 01 00 00]$.
This results in 
\begin{eqnarray}
\Phi^{(0, 1)}_{R^+} &=& \frac{1}{2 \eta^6(\tau) } 
(-2\theta_2(2\tau)\theta_3(2\tau)\theta_4^4(2\tau))=
-\frac{1}{2 \eta^{6} (\tau) }\theta_2^2(\tau)\theta_4^4(2\tau).
\end{eqnarray}
In the $\Phi^{(0,1)}_{NS^-}$ sector the characters which are present are 
$[00 00 00, 10 00 00] $ and  \\ $[11 00 00, 01 00 00]$  lead to 
\begin{eqnarray}
 \Phi^{(0,1)}_{NS^-} &=& \frac{1}{2 \eta^6(\tau)} (\theta_3^2(2\tau)-\theta_2^2(2\tau))\theta_4^4(2\tau), 
 \\ \nonumber
 &=& \frac{1}{2 \eta^6(\tau) } \theta_4^2(\tau)\theta_4^4(2\tau).
\end{eqnarray}
Finally the characters which survive the $g'$ insertion in 
$\Phi^{(0,1)}_{NS^-}$ are $-[00 00 00, 10 00 00]$ and  $[11 00 00, 01 00 00]$
giving rise to 
\begin{eqnarray}
 \Phi^{(0,1)}_{NS^+} &=& -  \frac{1}{2 \eta^{6}(\tau)  } 
 (\theta_3^2(2\tau)+ \theta_2^2(2\tau))\theta_4^4(2\tau), \\ \nonumber
 &=& -\frac{1}{2 \eta^{6}(\tau) } \theta_3^2(\tau)\theta_4^4(2\tau).
\end{eqnarray}
Now combining this along with the corresponding $D6$ characters as in 
(\ref{decompk3}) we obtain 
\begin{eqnarray}\label{z01}
 {\cal Z}_{\rm new}|_{(0, 1)} =-2 \frac{1}{\eta^{24} (\tau) } 
 \Gamma^{(0,1)}_{2,2} \times E_4\left[  - \frac{1}{2} ( {\cal E}_2(\tau) - 2{\cal E}_4(\tau) ).
 \right]  E_4
\end{eqnarray}
Here there we have used identities which relate the $\theta$ functions to Eisenstein series which 
are provided in the appendix. 
Using the action of  $g'^2$ which is given by  
\begin{equation} \label{gprimesqr}
(g')^2 =  \rho_L \left[ 
\left( \begin{matrix} 1 & 0\\ 0 & 1 \end{matrix} \right)
\left( \begin{matrix} 1 & 0\\ 0 & 1 \end{matrix} \right)
\left( \begin{matrix} -1 & 0\\ 0 & -1 \end{matrix} \right)
\left( \begin{matrix} -1 & 0\\ 0 &-1 \end{matrix} \right)
\left( \begin{matrix} -1 & 0\\ 0 & -1 \end{matrix} \right)
\left( \begin{matrix} -1 & 0\\ 0 & -1 \end{matrix} \right) \right] ,
\end{equation}
and the character list in table \ref{fznew} the 
contributions for the 
$\Phi^{(0, 2)}$ are evaluated. This results in 
\begin{eqnarray} \nonumber
 \Phi^{(0,2)}_{R^+} &=&-  \frac{1}{2 \eta^6(\tau)}
 4(\theta_2^5(2\tau)\theta_3(2\tau)+\theta_3^5(2\tau)\theta_2(2\tau))=- \frac{1}{2 \eta^6(\tau) } \theta_2^2(\theta_3^4+\theta_4^4), 
 \\ \nonumber
 \Phi^{(0,2)}_{NS^-} &=& \frac{1}{2\eta^6(\tau) }
  (\theta_3^6(2\tau)-\theta_2^6(2\tau)-3\theta_2^2(2\tau)\theta_3^4(2\tau)+3\theta_2^4(2\tau)\theta_3^2(2\tau)), \\ \label{gprm02}
  \Phi^{(0,2)}_{NS^+} &=&  \frac{1}{2\eta^6(\tau) }
   (-\theta_3^6(2\tau)-\theta_2^6(2\tau)-3\theta_2^2(2\tau)\theta_3^4(2\tau)-3\theta_2^4(2\tau)\theta_3^2(2\tau)).
\end{eqnarray}
Again combining these with the corresponding $D6$ characters 
and after using identities (\ref{int02id}) which relate the theta functions to 
Eisenstein series we obtain 
\begin{equation}\label{z02}
{\cal Z}_{\rm new}|_{(0, 2)} = -2 \frac{1}{\eta^{24}(\tau) }
\Gamma^{(0,2)}_{2,2} \times E_4\times  \left( - \frac{1}{6} E_6 + \frac{2}{3} {\cal E}_2 (\tau) E_4 
\right) .
\end{equation}
All the remaining terms in the new supersymmetric index can be 
obtained by performing modular transformations. 

On comparing the  coefficients of the twisted elliptic genus of the 
$2B$ orbifold given in (\ref{alphabeta}) with new supersymmetric index given in 
(\ref{z00}), (\ref{z01}), (\ref{z02})  we see that it agrees with the expression 
derived in (\ref{zneweis}) using general arguments for the standard embedding. 
It is important to realize that this agreement was due to non-trivial identities
relating  the theta functions  to  Eisenstein series together with the 
function ${\cal E}_2$ and ${\cal E}_4$.  Using the expression (\ref{nhnv}) 
we obtain $N_h -N_v = - 380$ for this model.

Now that we have shown the new supersymmetric index admits 
a decomposition in terms of the twisted elliptic genus for standard embeddings,  
the rest of the analysis in section \ref{thrcorr}  can be applied.  
Therefore we conclude that the difference in one loop gauge thresholds 
when the Wilson line is embedded in the unbroken $E_8$ is the theta lift 
of  twisted elliptic genus.

\section{Non-standard embeddings}\label{nonstdsec}

In this section we study the non-standard embeddings of heterotic compactifications
of $K3\times T^2$ orbifolded by $g'$ belonging to the conjugacy class $2A$. 
We first realize $K3$ as the $\mathbb{Z}_2$ orbifold of $T^4$ and consider the
$2$ non-standard embedding studied in \cite{Stieberger:1998yi}. 
We then move one to the situation in which $K3$ is realized as 
the $\mathbb{Z}_4$ orbifold of $T^4$ and $g'$ is implemented as  given in 
equations (\ref{orb1}) and (\ref{orb2}). 
We consider all the $12$ non-standard embeddings studied in \cite{Stieberger:1998yi}. 
In these orbifold limits, the various embeddings are implemented by 
different lattice shifts in the $E_8\times E_8$. 
From the spectrum of these embeddings we show that the 
they can be organized into $4$ types depending on the difference 
$N_h-N_v$ which take values $-12, 52, 84, 116$ for these types. 
The value $-12$ as we have seen corresponds to the standard type. 
The new supersymmetric index for all the embeddings  also  depends only  on 
$N_h-N_v$.  After turning on  the Wilson line  we show that the new supersymmetric 
index as well as the difference in one loop gauge thresholds depends on $N_h -N_v$ and
the instanton numbers of the embedding.

\subsection{Massless spectrum} \label{embed}

We can evaluate the massless spectrum of the non-standard embeddings 
 by following the same method as 
discussed in section \ref{standspec}. 
The spectrum for various non-standard embeddings
 of $K3\times T^2$  without the $g'$ orbifold were obtained   in \cite{Honecker:2006qz}. 
Essentially the orbifold by $g'$ changes the
degeneracy formula given in (\ref{unmod})   by changing the number of fixed points
of  the various twisted sectors as discussed around  (\ref{dgen2})   for the orbifold
in (\ref{orb1}), (\ref{orb2}).  
The various embeddings are determined by the lattice shifts in 
$E_8\times E_8$. 
In table \ref{z2spec}, we first tabulate the spectrum for embeddings when 
$K3$ is realized as the $T^4/\mathbb{Z}_2$ orbifold and 
$g'$ as half shift given by  following  orbifold actions
\begin{eqnarray}\label{z2chl}
g: \quad ( x_1, x_2, y_1, y_2, y_3, y_4) \sim  (x_1, x_2,-y_1,-y_2,-y_3,-y_4), \\ \nonumber
g': \quad ( x_1, x_2, y_1, y_2, y_3) \sim  (x_1+\pi, x_2, y_1+\pi, y_2, y_3, y_4).
\end{eqnarray}
The spectrum for the   $12$ non-standard embeddings 
when for  $K3$ is at  the $T^4/\mathbb{Z}_4$ orbifold limit 
 with $g'$ as shifts given in (\ref{orb2}) are listed 
in tables \ref{spectrumstand2},\ref{spectrumns1}
\ref{spectrumns2} and \ref{spectrumns3}. 
In these tables the shifts are denoted by $(\gamma; \tilde \gamma)$ where 
$\gamma, \tilde \gamma$ are $8$ dimensional vectors in $E_8\times E_8$. 
We observe from these tables that 
the the orbifold by $g'$ results in  only 
$4$ distinct values of $N_h-N_v$ given by 
$-12, 52, 84, 116$,  the value $-12$ corresponds to the
standard embedding. 
We classify these embeddings as type 0, type 1, type 2 and type 3 respectively.

\begin{table}[H]
\renewcommand{\arraystretch}{0.5}
\begin{center}
\vspace{-0.3cm}
\begin{tabular}{|c|c|c|}
\hline
Gauge group, Shift $(\gamma;\tilde{\gamma})$ & Sector & Matter \\ \hline
 & & \\
$E_7\times SU(2)\times E_8$& $g^0$ & $({\bf56;2})+4$({\bf1;1})\\ \cline{2-3}
 & & \\
$(1,-1,0^6;0^8)$  & $g^1$ & 4({\bf 56;1})+16({\bf 1;2}) \\ \hline
 & & \\
$ E_7\times SU(2)\times SO(16)$& $g^0$ & ({\bf 56,2;1})+4({\bf 1,1;1})\\
 & & +({\bf 1,1;128})\\ \cline{2-3}
 & & \\
$(1^2,0^6;2,0^7)$ &$g^1$ & 4({\bf 1,2;16})\\ \hline
\end{tabular}
\end{center}
\vspace{-0.5cm}
\caption{spectrum for different embeddings with $K_3$ as $T^4/Z_2$. 
The first shift realizes $N_h-N_v = -12$, while the second shift realizes 
$N_h -N_v = 116$. }\label{z2spec}
\renewcommand{\arraystretch}{0.5}
\end{table}

\begin{table}[H]
\renewcommand{\arraystretch}{0.5}
\begin{center}
\vspace{-0.5cm}
\begin{tabular}{|c|c|c|}
\hline
Gauge group, Shift $(\gamma;\tilde{\gamma})$ &Sector & Matter \\ \hline
 & & \\
$E_7\times U(1)\times E_8$ & $g^0$ & $({\bf56;1})+2({\bf1;1})$\\ \cline{2-3}
 & & \\
 & $g^1+g^3$ & $2({\bf 56;1})+4({\bf 1;1})+12({\bf 1;1})$\\ \cline{2-3}
 & & \\
$(1, 1,0^6;0^8)$ & $g^2$ & $3({\bf 56;1})+16({\bf 1;1})$\\ \hline
 & & \\
$E_7\times U(1)\times E_7\times SU(2)$ & $g^0$ & $ ({\bf 56;1,1})+2({\bf1;1,1})$\\ \cline{2-3}
 & & \\
 & $g^1+g^3$ & $6({\bf1;1,2})+2({\bf1;1,2})+2({\bf 1;56,1})$\\ \cline{2-3}
 & & \\
$(1,1,0^6;2,2,0^6)$ & $g^2$ & $1({\bf 56;1,1})+16({\bf1;1,1})$\\ \hline
 & & \\
$SO(12)\times SU(2)\times U(1)\times E_8$ & $g^0$ & $({\bf 12,2;1})+({\bf32,1;1})+2({\bf1,1;1})$\\ \cline{2-3}
 & & \\
 & $g^1+g^3$ & $6({\bf1,2;1})+4({\bf 12,1;1})$\\ 
$(3,1,0^6;0^8)$ & & $2({\bf 1,2;1})+2({\bf 32,1;1})$\\ \cline{2-3}
 & & \\
 & $g^2$ & $16({\bf1,1;1})+3({\bf 12,2;1})+({\bf 32,1;1})$\\ \hline
\end{tabular}
\end{center}
\vspace{-0.5cm}
\caption{Spectrum of 2A orbifold of $K3\times T^2$ for different embeddings 
belonging to type 0 for $K_3$ as $T^4/\mathbb{Z}_4$ with 
$N_h - N_v= -12$. }\label{spectrumstand2}
\renewcommand{\arraystretch}{0.5}
\end{table}

\begin{table}[H]
\renewcommand{\arraystretch}{0.5}
\begin{center}
\vspace{-0.5cm}
\begin{tabular}{|c|c|c|}
\hline
Gauge group, Shift $(\gamma;\tilde{\gamma})$ &Sector & Matter \\ \hline
 & & \\
$E_7\times U(1)\times SO(16)$ & $g^0$ & $({\bf 56;1})+2({\bf1;1})$\\ \cline{2-3}
 & & \\
 & $g^1+g^3$ & $8({\bf1;16})$\\ \cline{2-3}
 & & \\
$(1,1,0^6;4,0^7)$ & $g^2$ & $3({\bf56;1})+16({\bf1;1})$\\ \hline
 & & \\
$SO(12)\times SU(2)\times U(1)\times E_7\times SU(2)$ & $g^0$ & $({\bf12,2;1,1})+({\bf32,1;1,1})+2({\bf1,1;1,1})$\\ \cline{2-3}
 & & \\
 & $g^1+g^3$ & $4({\bf1,2;1,2})+2({\bf 12,1;1,2})$\\ \cline{2-3}
 & & \\
$(3,1,0^6;2,2,0^6)$ & $g^2$ & $16({\bf1,1;1,1})+({\bf 12,2;1,1})+3({\bf 32,1;1,1})$\\ \hline
 & & \\
$SO(12)\times SU(2)\times U(1)\times SO(16)$ & $g^0$ & $({\bf12,2;1})+({\bf 32,1;1})+2({\bf 1,1;1})$\\ \cline{2-3}
 & & \\
 & $g^1+g^3$ & $2({\bf 1,2;16})$\\ \cline{2-3}
 & & \\
$(3,1,0^6;4,0^7)$ & $g^2$ & $16({\bf 1,1;1})+3({\bf 12,2;1}0+({\bf 32,1;1})$\\ \hline
\end{tabular}
\end{center}
\vspace{-0.5cm}
\caption{Spectrum of 2A orbifold of $K3\times T^2$ for different embeddings in type 1
for $K_3$ as $T^4/\mathbb{Z}_4$ with  $N_h -N_v= 52$. }\label{spectrumns1}
\renewcommand{\arraystretch}{0.5}
\end{table}

\begin{table}[H]
\renewcommand{\arraystretch}{0.5}
\begin{center}
\vspace{-0.5cm}
\begin{tabular}{|c|c|c|}
\hline
Gauge group, Shift $(\gamma;\tilde{\gamma})$ &Sector & Matter \\ \hline
 & & \\
$E_7\times U(1)\times SU(8)\times U(1)$ & $g^0$ & $({\bf 56;1})+({\bf 1;8})+({\bf 1;56})+2({\bf 1;1})$\\ \cline{2-3}
 & & \\
 & $g^1+g^3$ & $6({\bf 1;1})+2({\bf1;1})+2({\bf 1;\bar{28}})$\\ 
 $(1,1,0^6;1^7,-1)$& & $+4({\bf 1,8})$\\ \cline{2-3}
 & & \\
 & $g^2$ & $6({\bf 1;8})+2({\bf 1;8})$\\ \hline
 & & \\
 & $g^0$ & $({\bf 27,2;1})+({\bf 1,2;1})+({\bf1,1;64})$\\
$E_6\times SU(2)\times U(1)\times SO(14)\times U(1)$& & $+2({\bf 1,1;1})$\\ \cline{2-3}
 & & \\
 & $g^1+g^3$ & $6({\bf 1,1;1})+4({\bf 1,2;1})$\\
$(2,1,1,0^5;2,0^7)$ & &$+2({\bf 27,1;1})+2({\bf 1,1;14 })$\\ \cline{2-3}
 & & \\
 & $g^2$ & $({\bf 1,2;14})+6({\bf1,2;1})$\\ \hline
\end{tabular}
\end{center}
\vspace{-0.5cm}
\caption{Spectrum of 2A orbifold of $K3\times T^2$ for different embeddings in type 2 for 
$K_3$ as $T^4/\mathbb{Z}_4$ with $N_h -N_v= 84$. }\label{spectrumns2}
\renewcommand{\arraystretch}{0.5}
\end{table}

\begin{table}[H]
\renewcommand{\arraystretch}{0.4}
\begin{center}
\vspace{-0.5cm}
\begin{tabular}{|c|c|c|}
\hline
Gauge group, Shift $(\gamma;\tilde{\gamma})$ &Sector & Matter \\ \hline
 & & \\
 & $g^0$ & $({\bf27,2;1,1})+({\bf1,2;1,1})+({\bf1,1;16,4})$\\
$E_6\times SU(2)\times U(1); SO(10)\times SO(6)$ & & $+2({\bf 1,1;1,1})$\\ \cline{2-3}
 & & \\
 & $g^1+g^3$ & $4({\bf1,1;1,4})+2({\bf1,2;1,4})$\\
 $(2,1,1,0^5;2^3,0^5)$ & & $+2({\bf1,1;\bar{16},1})$\\ \cline{2-3}
 & & \\
 & $g^2$ & $3({\bf1,2;10,1})+({\bf1,2;1,6})$\\ \hline
 & & \\
$SU(8)\times SU(2)\times SO(10)\times SO(6)$ & $g^0$ & $({\bf28,2;1,1})+({\bf1,1;16,4})+2({\bf1,1;1,1})$\\ \cline{2-3}
 & & \\
 & $g^1+g^3$ & $2({\bf8,1;1,4})$\\ \cline{2-3}
 & & \\
$(3,1^5,0^2;2^3,0^5)$ & $g^2$ & $16({\bf1,1;1})+3({\bf12,2;1,6})+({\bf1,2;10,1})$\\ \hline
 & & \\
$SU(8)\times SU(2)\times SO(14)\times U(1)$ & $g^0$ & $({\bf28,2;1})+({\bf1,1;64})+2({\bf1,1;1})$\\ \cline{2-3}
 & $g^1+g^3$ & $4({\bf\bar{8},1;1})+2({\bf8,2;1})$\\ \cline{2-3}
 & & \\
$(3,1^5,0^2;2,0^7)$ & $g^2$ & $3({\bf1,2;14})+2({\bf1,2;1})$\\ \hline
 & & \\
$SU(8)\times U(1)\times SO(12)\times SU(2)\times U(1)$ & $g^0$ & ${\bf(8;1,1)}+{\bf(56;1,1)}+{\bf(1;12,1)}$\\ 
& & $({\bf1;32,1})+2({\bf1;1,1})$\\ \cline{2-3}
 & & \\
$(1^7,-1;3,1,0)$ & $g^1+g^3$ & $4({\bf1;1,2})+2({\bf1;12,1})+2({\bf8;1,2})$\\ \cline{2-3}
 & & \\
 & $g^2$ & $6({\bf8;1,1})+2({\bf8;1,1})$\\ \hline
 & & \\
\end{tabular}
\end{center}
\vspace{-0.5cm}
\caption{Spectrum of 2A orbifold of $K3\times T^2$ for different embeddings in type 3 
for $K_3$ as $T^4/\mathbb{Z}_4$ with $N_h -N_v = 116$.}\label{spectrumns3}
\renewcommand{\arraystretch}{0.4}
\end{table}

\noindent
Finally in table \ref{org1} and \ref{org2} we group the shifts according to the type based on 
the value of $N_h-N_v$.

\begin{table}[H]
\renewcommand{\arraystretch}{1.0}
\begin{center}
\vspace{-0.3cm}
\begin{tabular}{|cc|c|c|}
\hline
$\gamma$ & $\tilde{\gamma}$ & Type & $N_h-N_v$\\
\hline
(1,1,0,0,0,0,0,0) & (0,0,0,0,0,0,0,0)& Type 0 & -12 \\
(1,-1,0,0,0,0,0,0) & (2,0,0,0,0,0,0,0)& Type 3& 116 \\
\hline
\end{tabular}
\end{center}
\vspace{-0.5cm}
\caption{Lattice shifts in  the $2A$ orbifold with   $K3=T^4/\mathbb{Z}_2$ and  $N_h-N_v$}\label{org1}
\renewcommand{\arraystretch}{1.0}
\end{table}

\begin{table}[H]
\renewcommand{\arraystretch}{1.0}
\begin{center}
\vspace{0.5cm}
\begin{tabular}{|cc|c|c|}
\hline
$\gamma$ & $\tilde{\gamma}$ & Type & $N_h-N_v$\\
\hline
(1,-1,0,0,0,0,0,0) & (0,0,0,0,0,0,0,0)& & \\
(1,1,0,0,0,0,0,0) & (0,0,0,0,0,0,0,0)& Type 0 &-12\\
(1,1,0,0,0,0,0,0)& (2,2,0,0,0,0,0,0)& & \\
(3,1,0,0,0,0,0,0) & (0,0,0,0,0,0,0,0)& & \\
\hline
(1,1,0,0,0,0,0,0)& (4,0,0,0,0,0,0,0)& & \\
(3,1,0,0,0,0,0,0)& (4,0,0,0,0,0,0,0)& Type 1 &52\\
(3,1,0,0,0,0,0,0)& (2,2,0,0,0,0,0,0)& & \\
\hline
(2,1,1,0,0,0,0,0)& (2,0,0,0,0,0,0,0)& & \\
(1,1,0,0,0,0,0,0)& (1,1,1,1,1,1,1,-1)& Type 2 &84\\
\hline
(2,1,1,0,0,0,0,0) & (2,2,2,0,0,0,0,0)& & \\
(3,1,1,1,1,1,0,0) & (2,0,0,0,0,0,0,0)& Type 3 &116\\
(3,1,1,1,1,1,0,0) & (2,2,2,0,0,0,0,0)& & \\
(1,1,1,1,1,1,-1) & (3,1,0,0,0,0,0,0)& & \\
\hline
\end{tabular}
\end{center}
\vspace{-0.5cm}
\caption{Lattice shifts in the $2A$ orbifold  with $K3 = T^4/\mathbb{Z}_4$ and $N_h-N_v$}\label{org2}
\renewcommand{\arraystretch}{1.0}
\end{table}

\subsection{ New supersymmetric index}

In this section we evaluate the new supersymmetric index for all the 
embeddings discussed in section \ref{embed}. 
We will show that for the when the Wilson line is not turned on, the  index ${\cal Z}_{\rm new}$
for the $2A$ orbifold of $K3\times T^2$ depends only on the  $4$ types of the 
lattice shifts organized  in tables (\ref{org1}) and (\ref{org2}). ${\cal Z}_{\rm new}$
is invariant for  any lattice shift belonging to a given type.  
When the Wilson line is turned on, then the index depends both on the
type as well as the instanton number corresponding to the  lattice shift.

Let us first discuss the case without the Wilson line. 
Evaluating the trace defined in (\ref{znew}) we see that 
it reduces to 
 \begin{eqnarray} \label{z4nonstd}
{\cal Z}_{\rm new}(q, \bar q)  = - \frac{1}{2\eta^{20}( \tau)  }\sum_{a, b =0}^{\nu-1} \sum_{r, s =0}^1 
e^{- \frac{2\pi i a b}{\nu^2}} 
Z_{E_8 }^{(a, b) } (\tau) \times Z_{E_8' }^{(a, b) } (\tau) 
 \times \frac{1}{2 \nu } F( a, r,  b,  s; q) \Gamma^{(r, s)}_{2, 2} (q, \bar q), \nonumber \\
\end{eqnarray}
where $\nu = 2, 4$ depending on the whether $K3$ is realized as
a $T^4/\mathbb{Z}_2$ or $T^4/\mathbb{Z}_4$ orbifold. 
The partition function over the shifted $E_8$ lattices are defined by 
\begin{eqnarray}
Z_{E_8}^{a,b}(q)&=&\frac{1}{2}\sum_{\al,\beta=0}^{1}
e^{-i\pi\beta\frac{a}{\nu}\sum_{I=1}^8\gamma^{I}}
\prod_{I=1}^8\theta\left[
\begin{smallmatrix}\al+2\frac{a}{\nu}\gamma^I \\ \beta+2\frac{b}{\nu}\gamma^I\end{smallmatrix}\right],\\ 
Z_{E_8'}^{a,b}(q)&=&\frac{1}{2}\sum_{\al,\beta=0}^{1}e^{-i\pi\beta\frac{a}{\nu}
\sum_{I=1}^8\tilde{\gamma}^{I}}\prod_{I=1}^8\theta\left[
\begin{smallmatrix}\al+2\frac{a}{\nu}\tilde{\gamma}^I \\ \beta+2\frac{b}{\nu}\tilde{\gamma}^I
\end{smallmatrix}\right],
\end{eqnarray}
where $\gamma, \tilde\gamma$ are the shifts in the two $E_8$ lattices. 
The trace over the $T^4$ directions  is as defined in (\ref{fabrs}). 
However the $g, g'$ correspond to the actions in (\ref{z2chl}) for the $\mathbb{Z}_2$ orbifold 
limit of $K3$ and to actions (\ref{orb1}) and (\ref{orb2})  for the $\mathbb{Z}_4$ orbifold
limit of $K3$. 
This trace is given by 
\begin{eqnarray}
F ( a, r, b, s; q)  =
  k^{(a,r,b,s)}_{(\nu)} \eta^2(\tau)q^{\frac{-a^2}{\nu^2}}\frac{1}{\theta_1 ^2(\frac{a\tau+b}{\nu}, \tau )}\; ,
\end{eqnarray}
where  the $k$'s are read out from the following matrices. 
\begin{eqnarray}
k^{(a,0,b,0)}_{(2)} &=& 
64\left ( \begin{matrix}
0&1\\
1 &e^{-\pi i(2-\Gamma^2)/4}
 \end{matrix} \right), 
 \qquad 
 k^{(a,0,b,1)}_{(2)} = 
64\left ( \begin{matrix}
0&1\\
0 & 0 
 \end{matrix} \right), 
  \\ \nonumber
 k^{(a,1,b,0)}_{(2)}&=& 
64\left ( \begin{matrix}
0&0\\
1 & 0
 \end{matrix} \right), 
 \qquad 
 k^{(a,1,b,1)}_{(2)}=
64\left ( \begin{matrix}
0&0\\
0 &e^{-\pi i(2-\Gamma^2)/4}
 \end{matrix} \right),  \\ \nonumber
k^{(a,0,b,0)}_{(4)}&=&
16\left(\begin{matrix} 0 & 1 & 4 & 1\\ 1 & e^{-\pi i\frac{1}{16}(2-\Gamma^2)}& e^{-\pi i\frac{1}{8}(2-\Gamma^2)} & e^{-\pi i\frac{3}{16}(2-\Gamma^2)}\\4 & e^{\pi i\frac{3}{8}(2-\Gamma^2)} & 4e^{-\pi i\frac{1}{4}(2-\Gamma^2)} & e^{\pi i\frac{1}{8}(2-\Gamma^2)}\\1 & e^{\pi i\frac{9}{16}(2-\Gamma^2)} & e^{\pi i\frac{1}{8}(2-\Gamma^2)} & e^{-\pi i\frac{9}{16}(2-\Gamma^2)}
\end{matrix} \right), \\ \nn
 k^{(a,0,b,1)}_{(4)}&=&
 16\left(\begin{matrix} 0 & 1 & 4 & 1\\ 0 & 0 & 0 & 0\\0 & e^{\pi i\frac{3}{8}(2-\Gamma^2)} & 0 & e^{\pi i\frac{3}{8}(2-\Gamma^2)}\\0 & 0 & 0 & 0 \end{matrix} \right),\\  \nn
k^{(a,1,b,0)}_{(4)}&=&
16\left(\begin{matrix} 0  & 0 & 0 & 0\\ 1\quad & 0 & e^{-\pi i\frac{1}{8}(2-\Gamma^2)} & 0\\4   & 0 & 0 & 0\\1  & 0 & e^{\pi i\frac{1}{8}(2-\Gamma^2)} & 0
\end{matrix} \right), \\ \nn
 k^{(a,1,b,1)}_{(4)}&=&
 16\left(\begin{matrix} 0 & 0 & 0 & 0\\ 0 & e^{-\pi i\frac{1}{16}(2-\Gamma^2)} & 0 & e^{-\pi i\frac{3}{16}(2-\Gamma^2)}\\0 & 0 &4 e^{-\pi i\frac{1}{4}(2-\Gamma^2)} & 0\\0 & e^{\pi i\frac{9}{16}(2-\Gamma^2)} & 0 & e^{-\pi i\frac{9}{16}(2-\Gamma^2)}
\end{matrix} \right),
\end{eqnarray}
where $\Gamma^2 = \gamma^2 + \tilde\gamma^2$.
Using all this inputs we evaluate the new supersymmetric index for 
the  list of lattice shifts given in tables \ref{org1} and \ref{org2}.  This results in 
following general result 

\begin{eqnarray}\label{newindnon}
\cal{Z}_{\rm new}&=&-\frac{1}{\eta^{24}}\left\{ 2 \Gamma^{(0,0)}_{2, 2} 
E_4E_6  \right. \\ \nonumber
& & +\Gamma^{(0,1)}_{2,2} 
\left[(E_6+2{\cal E}_2(\tau) E_4 ) \left(\hat{b}{\cal E}_2^2(\tau)+ (\frac{2}{3} - \hat b)  E_4\right)\right]\\ \nn
&&+\Gamma^{(1,0)}_{2,2}
\left [\left(E_6-{\cal E}_2(\frac{\tau}{2}) E_4\right)\left({\frac{\hat b}{4}}{\cal E}_2^2(\frac{\tau}{2} )+ 
( \frac{2}{3} - \hat b) E_4\right) \right]\\ \nn
&&\left. +\Gamma^{(1,1)}_{2,2}
\left[ \left(E_6-{\cal E}_2(\frac{\tau+1}{2})E_4\right) 
\left({\frac{\hat b}{4}}{\cal E}_2^2(\frac{\tau+1}{2})+ ( \frac{2}{3} - \hat b) E_4\right) \right] \right\} .
\end{eqnarray}
The value of $\hat b$ for each of type of embeddings is given in table \ref{bvalue}. 
 \begin{table}[H]
\renewcommand{\arraystretch}{1.0}
\begin{center}
\vspace{0.5cm}
\begin{tabular}{|c|c|c|c|c|}
\hline
Type & Type 0  & Type 1 &Type 2 &Type 3 \\ \hline
$\hat b$  & 0 & $\frac{4}{9}$ & $\frac{2}{3}$ & $\frac{8}{9}$\\
\hline 
\end{tabular}
\end{center}
\vspace{-0.5cm}
\caption{Value of $\hat b$ for each type of lattice shift }\label{bvalue}
\renewcommand{\arraystretch}{1.0}
\end{table}
Thus the values $\hat b$ takes are discrete and just depends on the type of embedding or
lattice shift.  
In fact since $N_h - N_v$ remains constant in each type of embedding 
we can relate it to $\hat b$. 
This relation can be found by using the equation  in (\ref{nhnv}) and is given by 
\begin{equation}\label{bnhnv}
N_h-N_v=144\hat b-12.
\end{equation}
Note that  standard embedding belongs the case $\hat b=0$, also note
that the only non-standard embedding of the $2A$ orbifold when  $K3$ 
is realized as $T^4/\mathbb{Z}_2$  as seen in table \ref{org1}  belongs 
to type 3.  One important point to  emphasize is that the new supersymmetric
index in (\ref{newindnon}) still can be decomposed in terms of the 
twisted elliptic genus of $K3$. Comparing (\ref{zneweis}) for the $2A$ orbifold
with (\ref{newindnon})
the only difference is that the lattice sum $E_4$ has been replaced by 
$ \left(\hat{b}{\cal E}_2^2(\tau)+ (\frac{2}{3} - \hat b)  E_4\right)$ for the $(0,1)$
sector. 
The lattice sum $\left(E_6-{\cal E}_2(\frac{\tau}{2}) E_4\right)$ associated by the 
$2A$ orbifold remains the same. Similar statements can be 
made for all the other sectors.

Let us now turn on the Wilson line in the $E_8'$ lattice and evaluate the 
new supersymmetric index. 
To do this we follow the procedure in \cite{Stieberger:1998yi}. 
First the partition function in the $E_8'$ lattice is evaluated with  
a chemical potential along one of $U(1)$ directions. 
The lattice sum then becomes

\begin{eqnarray}
Z_{E_8'}^{a,b}(\tau,z)=\frac{1}{2}\sum_{\al,{\beta}=0}^{1}e^{-i\pi\beta\frac{a}{4}\sum_{I=1}^8\tilde{\gamma}^{I}}\prod_{I=1}^6\theta\left[\begin{smallmatrix}\al+2\frac{a}{4}\tilde{\gamma}^I
 \\ {\beta}+2\frac{b}{4}\tilde{\gamma}^I\end{smallmatrix}\right](\tau)\prod_{I=7}^8\theta\left[\begin{smallmatrix}\al+2\frac{a}{4}\tilde{\gamma}^I \\ {\beta}+2\frac{b}{4}\tilde{\gamma}^I\end{smallmatrix}\right](\tau,z).
\end{eqnarray}
This modified lattice sum  $Z_{E_8'}^{a,b}(\tau,z)$ is then coupled to the
$\Gamma_{3, 2}$ lattice using the $\otimes$ product defined in the appendix. 
It was shown  in \cite{Stieberger:1998yi}
that for all orbifold realizations of $K3$, the new supersymmetric index 
just depends on instanton numbers of the embedding or the lattice shifts. 
The result is given by the expression

\begin{equation}
{\cal Z}_{\rm new}=-\frac{1}{6\eta^{24} }\Gamma_{3,2}(q,\bar{q}) \otimes 
[n_1 E_{4, 1}  E_6 +n_2 E_{6, 1} E_{4}],
\end{equation}
where $n_1, n_2$ are the instanton numbers of the embedding and $n_1 + n_2 = 24$. 
For the standard embedding $n_1 = 24, n_2 =0$. 
Thus the new supersymmetric index with the Wilson line is sensitive to 
the the instanton numbers. 

For compactifications on $(K3\times T^2)/g'$  with $K3$ realized either by  $T^4/\mathbb{Z}_2$ or 
the $T^4/\mathbb{Z}_4$ and  $g'$ in the $2A$ conjugacy class, the new 
supersymmetric index with the Wilson line depends on $\hat b$ which is 
related to $N_h-N_v$ of the model by (\ref{bnhnv}) and also the instanton number 
of the embedding. 
The result for the index for all the embeddings  can be summarized 
in the following compact expression

\begin{eqnarray} \label{abc}
& & {\cal Z}_{\rm new} 
= - \frac{1}{ \eta^{24}} \left\{ \Gamma^{(0,0)}_{3, 2}  \otimes \frac{1}{12}
[n_1 E_{4, 1}  E_6 +n_2 E_{6, 1} E_{4}] 
\right. \\ \nonumber
& &   + \Gamma^{(0,1)}_{3, 2} \otimes  \left[
\hat a E_{4,1}(E_6+2{\cal E}_2(\tau) E_4)+
\hat b {\cal E}_2(\tau)^2 (E_{6,1}+2 {\cal E}_2(\tau)E_{4,1})+
\hat c E_{4}(E_{6,1}+2 {\cal E}_2(\tau)E_{4,1}) \right]  \\ \nonumber
& &\left. +\Gamma^{(1,0)}_{3, 2} \otimes [ \quad \cdot \quad] 
 +\Gamma^{(1,1)}_{3, 2} \otimes [ \quad \cdot \quad] 
 \right\}.
\end{eqnarray}
Here the parameters  $\hat a , \hat c$  depend on the instanton numbers 
$n_1, n_2$ of the embedding and the value of $\hat b$ by 
\begin{eqnarray} \label{eqnac}
\hat a=\frac{n_1}{36}- \frac{\hat b}{2}, \qquad
\hat c=  \frac{2}{3} - \hat a - \hat b \;.
\end{eqnarray}
The  $ [ \quad \cdot \quad] $ denotes the corresponding term
obtained by modular transformation of the $(0,1)$ sector. 
For example in the $(1, 0)$ sector, we replace the terms with 
${\cal E}_2(\tau)$ of the $(0, 1)$ sector to $-\frac{1}{2}{\cal E}_2(\frac{\tau}{2})$. 
Similarly in the $(1,1)$ we have $-\frac{1}{2}{\cal E}_2(\frac{\tau+1}{2} )$. 
We summarize the values of $\hat a, \hat b, n_1$
 for each of the shifts considered 
in  the tables \ref{tblz21} and \ref{tbl1}. 
Using these tables and equation (\ref{abc}), the result for the new supersymmetric
index with the Wilson line for these orbifolds can be read out. 

\begin{table}[H]
\renewcommand{\arraystretch}{1.0}
\begin{center}
\vspace{0.5cm}
\begin{tabular}{|c|cc|c|ccc|}
\hline
Type & $\gamma$ &$\tilde{\gamma}$& $(n_1, n_2)$&$\hat a$ &$\hat b$ &$\hat c$\\
\hline
Type 0 & (1,-1,0,0,0,0,0,0) & (0,0,0,0,0,0,0,0) & (24,0) & 2/3 & 0 & 0 \\
\hline
Type 3 & (1,-1,0,0,0,0,0,0) & (2,0,0,0,0,0,0,0) & (8,16) & -2/9  & 8/9  & 0 \\
\hline
\end{tabular}
\end{center}
\vspace{-0.1cm}
\caption{Lattice shifts for   $((T^4/\mathbb{Z}_2) \times T^2)/g'$ and their $\hat a, \hat b, \hat c$ values} \label{tblz21}
\renewcommand{\arraystretch}{1.0}
\end{table}

\begin{table}[H]
\renewcommand{\arraystretch}{1.0}
\begin{center}
\vspace{0.5cm}
\begin{tabular}{|c|cc|c|ccc|}
\hline
Type & $\gamma$ &$\tilde{\gamma}$& $(n_1, n_2)$&$\hat a$ &$\hat b$ &$\hat c$\\
\hline
Type 0 & (1,-1,0,0,0,0,0,0) & (0,0,0,0,0,0,0,0) & (24,0)& 2/3 &0&0\\
& (1,1,0,0,0,0,0,0) & (0,0,0,0,0,0,0,0) & (24,0)& 2/3 &0&0\\
& (3,1,0,0,0,0,0,0) & (0,0,0,0,0,0,0,0) & (24,0)& 2/3 &0&0\\
& (1,1,0,0,0,0,0,0) & (2,2,0,0,0,0,0,0) & (12,12)& 1/3 &0&1/3\\
\hline 
Type 1 & (1,1,0,0,0,0,0,0) & (4,0,0,0,0,0,0,0)&(16,8) &2/9 &4/9 &0\\
& (3,1,0,0,0,0,0,0) & (4,0,0,0,0,0,0,0)&(16,8) &2/9 &4/9 &0\\
& (3,1,0,0,0,0,0,0) & (2,2,0,0,0,0,0,0)&(20,4)&1/3&4/9&-1/9 \\
\hline
Type 2 & (2,1,1,0,0,0,0,0) & (2,0,0,0,0,0,0,0)&(12,12)&0&2/3 &0 \\
& (1,1,0,0,0,0,0,0) & (1,1,1,1,1,1,1,-1)&(6,18) &-1/6&2/3 &1/6\\
\hline 
Type 3& (2,1,1,0,0,0,0,0) & (2,2,2,0,0,0,0,0)&(12,12)&-2/9&8/9 &0\\
& (3,1,0,0,0,0,0,0) & (1,1,1,1,1,1,1,-1)&(14,10)&-1/18 &8/9 &-1/6\\
& (3,1,1,1,1,1,0,0) & (2,0,0,0,0,0,0,0)&(12,12)&-1/9&8/9 &-1/9 \\
& (3,1,1,1,1,1,0,0) & (2,2,2,0,0,0,0,0)&(12,12)&-1/9&8/9 &-1/9\\
\hline
\end{tabular}
\end{center}
\vspace{-0.1cm}
\caption{Lattice shifts for  $((T^4/\mathbb{Z}_4) \times T^2)/g'$ and their $\hat a, \hat b, \hat c$ values}
\label{tbl1}
\renewcommand{\arraystretch}{1.0}
\end{table}

\subsection{Difference of one loop gauge thresholds}

We now evaluate the difference in one loop gauge thresholds 
for all models whose new supersymmetric index is given by  (\ref{abc}). 
The one loop threshold for the group $G$ is given by (\ref{gthresh}). 
We take the $G$ to be the group the Wilson line is embedded in. 
Then  using (\ref{abc}) we obtain 
\begin{eqnarray}
 {\cal B}_G &=& - \frac{1}{ \eta^{24}} \left\{ 
 \Gamma_{3, 2}^{(0, 0 )}\otimes \frac{1}{ 288} [ n_1 ( \tew E_{4, 1} - E_{6, 1})  E_6
 + n_2( \tew E_{6, 1} - E_{4, 1} E_4 ) E_6  ]  \right.  \nonumber \\   \nonumber
&& +  \Gamma_{3, 2}^{(0, 1 )}\otimes \left [
\frac{\hat a}{24} (E_{4,1}\tew-E_{6,1})(E_6+2{\cal E}_2(\tau) E_4)  \right. \\ \nonumber
& &  +\frac{\hat c}{24} E_4\left(E_{6,1}\tew-E_{4,1}E_4 + 
2{\cal E}_2(\tau)(E_{4,1}\tew-E_{6,1})\right) \\ \nonumber
& &\left.  +\frac{\hat b}{120} \left(( E_4+4E_4(2\tau))(E_{6,1}\tew-E_{4,1}E_4  
 +2{\cal E}_2(\tau)E_{4,1}\tew-2{\cal E}_2(\tau)E_{6,1}\right) \right] \\ 
 & & 
 \left. + \Gamma_{3, 2}^{(1, 0  )}\otimes [\quad \cdot \quad ]  +   \Gamma_{3, 2}^{(1, 1  )}\otimes 
 [\quad \cdot \quad ]  \right\} .
\end{eqnarray}
where the terms in the $[\quad\cdot \quad]$ can be obtained by modular 
transformation from the corresponding term in the $(0, 1)$  sector. 
Note that we have used the identity 
\begin{equation}\label{multip}
 {\cal E}_2^2(\tau) = \frac{1}{5} \left( 4 E_4 (2\tau) + E_4  \right) ,
\end{equation}
in the terms proportional to $\hat b$. 
Similarly the terms for the gauge group $G'$ we obtain 
\begin{eqnarray}
 {\cal B}_{G'} &=& - \frac{1}{ \eta^{24}} \left\{  
 \Gamma_{3, 2}^{(0, 0 )}\otimes \frac{1}{ 288} [ n_1 E_{4, 1} ( \tew E_6 - E_4^2 ) + 
 n_2 ( \tew E_4 - E_6) ] \right.   \\ \nonumber
 & & +  \Gamma_{3, 2}^{(0, 1 )}\otimes \left [
 \frac{\hat a}{24} E_{4,1}(E_6\tew-E_4^2+2{\cal E}_2(\tau)(E_4\tew-E_6))+
 \frac{\hat c}{24}(E_4\tew-E_6)(E_{6,1}+2{\cal E}_2(\tau) E_{4,1}) \right. \\ \nn
&&\left. 
+\frac{\hat b}{120}(\tew E_4-E_6+8(\tew(2\tau)E_4(2\tau)-E_6(2\tau))(E_{6,1}+2{\cal E}_2(\tau) E_{4,1}) \right]
\\ \nonumber
& & \left. + \Gamma_{3, 2}^{(1, 0  )}\otimes [\quad \cdot \quad ]  +   \Gamma_{3, 2}^{(1, 1  )}\otimes 
 [\quad \cdot \quad ]  \right\} .
\end{eqnarray}
We now evaluate the difference in the threshold integrals. To simplify the expressions we use the 
following identities
\begin{eqnarray}
{\cal E}_2(\tau)=2\tew(2\tau)-\tew, \qquad 
E_6(2\tau)=\frac{{\cal E}_2(\tau)}{8}(11 {\cal E}_2^2(\tau)-3E_4),
\end{eqnarray} 
together with  (\ref{multip}) and 
\begin{equation}\label{cuberel}
{\cal E}_2(\tau)^3=\frac{3}{4}E_4{\cal E}_2(\tau)+\frac{1}{4}E_6 \, .
\end{equation}
 This results in the following 
expression for the threshold integral
\begin{eqnarray}\label{bbprimew}
& & \Delta_G(T, U, V) - \Delta_{G'} ( T, U, V) = \int_{\cal{F}} \frac{d^2\tau}{\tau_2} 
\{ {\cal B}_G - {\cal B}_{G'} \}  \\ \nonumber
\qquad &=&  \int_{\cal{F}} \frac{d^2\tau}{\tau_2}  \left\{ 
\Gamma^{(0,0)} \otimes 2 (n_2-n_1)A(z) 
\right. \\ \nn
\qquad &&- \Gamma^{0,1} \otimes 
\left[24A(z)(\frac{n_1-12}{18})-12B(z){\cal E}_2(\tau)(\frac{2}{3}-\frac{\hat b}{2}) 
 \right] \\ \nn
\qquad &&-\Gamma^{(1,0)}\otimes \left[
24A(z)(\frac{n_1-12}{18})+6B(z){\cal E}_2(\frac{\tau}{2} )(\frac{2}{3}-\frac{\hat b}{2})\right]
\\ \nn
\qquad &&\left. -\Gamma^{(1,1)}\otimes \left[ 
24A(z)(\frac{n_1-12}{18})+6B(z){\cal E}_2(\frac{\tau+1}{2})(\frac{2}{3}-\frac{\hat b}{2})\right]
\right\},
\end{eqnarray}
where  we have used the relations (\ref{ABrel}). Note that the 
integrands for all the  embeddings in table (\ref{tblz21}) and (\ref{tbl1}) just 
depend on the instanton number and the $\hat b$ which is related to the 
difference $N_h-N_v$.  One simple check of our result is that 
on setting $b=0, n_1 = 24$, the equation in (\ref{bbprimew}) reduces to 
the standard embedding result for the $2A$ orbifold of $K3$. 

The threshold integral  in (\ref{bbprimew}) 
over the fundamental domain can be performed using the 
methods  developed in \cite{justin3}. 
The details are provided in the appendix \ref{integrals}. 
Here we quote the final result. 
\begin{eqnarray} \label{findifthr}
 \Delta_G(T, U, V) - \Delta_{G'} ( T, U, V) & =&  48
 \left(  (  \frac{1}{2} - \frac{3\hat b}{8} ) \log(\rm{det}({\rm{Im}}(\Omega))^6\left|\Phi_6(U,T,V) \right|^2  )   
 \right. \\ \nonumber
& &  +  ( \frac{n_1}{72} - \frac{1}{3} + \frac{\hat b}{8} ) 
 \log(\rm{det}
\left(\rm{Im}(\Omega))^{10}\left| \Phi_{10}(U,T,V) \right|^2 \right) \\ \nonumber
 && \left. +   ( \frac{n_1}{72} - \frac{1}{3} + \frac{\hat b}{8} ) \log(\rm{det}
\left(\rm{Im}(\Omega))^{10}\left| \Phi_{10}(2U,T/2,V) \right|^2 \right) 
 \right)
\end{eqnarray}
Here $\Phi_{10}$ is the unique cusp form of weight $10$ under $Sp(2, \mathbb{Z})$, 
while $\Phi_{6}$ is the Siegel modular form of weight $6$ which is 
obtained from the theta lift of the elliptic genus of $K3$ twisted by the 
$2A$ orbifold action.  $\Phi_{6}$  was first constructed  as a theta lift  
in \cite{justin1}.  As expected for the standard embedding $ \hat b =0, n_1 =24$ 
the threshold integral reduces to only $\Phi_6$.

\section{Conclusions}\label{conclus}

We have explored ${\cal N}=2$ compactifications of heterotic string theory 
on orbifolds of $K3\times T^2$ by $g'$ which acts as 
a $\mathbb{Z}_N$ automorphism on $K3$ together with a 
$1/N$ shift on one of the circles of $T^2$. 
$g'$ can correspond to any of the $26$ conjugacy classes of 
the Mathieu group $M_{24}$. 
We showed that for the  standard embedding of the spin connection 
in one of the $E_8$ the new supersymmetric index can 
be written in terms of the elliptic genus of $K3$ twisted by $g'$. 
The difference in gauge thresholds are shown to be 
theta lifts of the twisted elliptic genus of these compactifications. 
This generalizes the observation in \cite{Datta:2015hza} as well as 
\cite{Angelantonj:2014dia,Angelantonj:2015nfa} who 
observed similar behaviour for non-supersymmetric compactifications
\footnote{In the case of non-supersymmetric compactifications, the 
difference in  the gauge threshold integrand  was the lattice sum  $\Gamma_{2,2}$ 
folded with 
a holomorphic function which resembled an index. }
We demonstrated this by explicitly studying 2 examples. The first one 
considered the $2A$ orbifold of $K3$  when $K3$ is realized as $T^4/\mathbb{Z}_4$. 
The result is same as that obtained in \cite{Datta:2015hza} where the $2A$ orbifold of 
$K3$ is obtained by taking $K3$ to be $T^4/\mathbb{Z}_2$. 
We also studied  the recently constructed \cite{Gaberdiel:2013psa} $2B$ orbifold of $K3$ 
when $K3$ is realized as ${\rm su}(2)^6$   rational conformal field theory.  
Finally we considered non-standard embeddings  for the  $2A$ orbifold of $K3$
and showed that the new supersymmetric index depends only on 
the difference $N_h-N_v$ of the model and the gauge threshold 
correction depends on the instanton number of the embedding as 
well as $N_h -N_v$.  The detailed spectrum of these compactifications 
has also be obtained. 

There are a number of directions which are worth exploring. 
One is to generalize the study of non-standard embedding to 
all the orbifold limits of $K3$, here we considered 
only  the limits $T^4/\mathbb{Z}_2$ and $T^4/\mathbb{Z}_4$ .  
Another direction is to study the type II duals of these theories. 
Not only this will teach us more about $S$-duality, but it will 
also involve the study of new Calabi-Yau manifolds. 
However perhaps the most  interesting extrapolation of the observations
of this paper 
is the fact that it is also possible to  consider compactifications of string theory
of type II  on $(K3\times T^2)/g'$ 
where $g'$ corresponds to any of the $26$ conjugacy classes of 
$M_{24}$. These compactifications preserve 
${\cal N}=4$ supersymmetry. The   theta lifts of the
twisted elliptic genus for all these cases should 
capture degeneracies of $1/4$ BPS dyons. 
The case of $g'$ in the conjugacy class $pA,\, p =1, 2, 3, 5, 7$ 
was studied in 
\cite{Dijkgraaf:1996it,LopesCardoso:2004xf,Shih:2005uc,Gaiotto:2005hc,Jatkar:2005bh,
justin1,David:2006yn,Dabholkar:2006xa}.  It will be 
certainly interesting to generalize the results regarding dyon partition functions
 to all the conjugacy 
classes of $M_{24}$. This will possibly 
 will teach us about black hole degeneracies  in ${\cal N}=4$ string 
theory and its relation to the symmetry $M_{24}$.

\acknowledgments 
We thank Ashoke Sen for useful discussions. 
The work of 
 A.C is funded by the CSIR JRF fellowship  09/079(2649)/2015-EMR-I.

\appendix
\section{Notations,  conventions and identities}\label{notation}
In this appendix we summarize the notations and conventions and properties of the modular 
functions used in this paper.
We define the generalized form of Jacobi theta functions as
\be \label{genjacfn}
\theta\left[\begin{smallmatrix}a\\ b\end{smallmatrix}\right](q,z)
=\sum_{k\in \mathbb{Z}}q^{\frac{1}{2}(k+\frac{a}{2})^2}e^{\pi i (k+\frac{a}{2})b}e^{(2\pi i z)(k+\frac{a}{2})}.
\ee
If the  variable $z$  is not stated in the argument  then it is understood to be the theta function is at $z=0$.
We  use $q = e^{2\pi i \tau}$ and $\tau$ interchangeably in the arguments of the 
modular functions. We also define 
\bea
\theta_1(\tau,z)=\theta\left[\begin{smallmatrix}1\\ 1\end{smallmatrix}\right](\tau,z) \qquad \theta_2(\tau,z)=\theta\left[\begin{smallmatrix}1\\ 0\end{smallmatrix}\right](\tau,z),\\ \nn
\theta_3(\tau,z)=\theta\left[\begin{smallmatrix}0\\ 0\end{smallmatrix}\right](\tau,z) \qquad \theta_4(\tau,z)=\theta\left[\begin{smallmatrix}0\\ 1\end{smallmatrix}\right](\tau,z).
\eea

In various manipulations the following Riemann bi-linear identities are useful
\bea \label{doublez}
\theta_1 ^2(\tau,z)&=&\theta_2(2\tau)\theta_3(2\tau,2z)-\theta_3(2\tau)\theta_2(2\tau,2z),\\ \nn
\theta_2 ^2(\tau,z)&=&\theta_2(2\tau)\theta_3(2\tau,2z)+\theta_3(2\tau)\theta_2(2\tau,2z),\\ \nn
\theta_3 ^2(\tau,z)&=&\theta_3(2\tau)\theta_3(2\tau,2z)+\theta_2(2\tau)\theta_2(2\tau,2z),\\ \nn
\theta_4 ^2(\tau,z)&=&\theta_3(2\tau)\theta_3(2\tau,2z)-\theta_2(2\tau)\theta_2(2\tau,2z).\\ \nn
\eea
At $z=0$, these identities reduce to 
\bea \label{double}
& & \theta_2^2=2\theta_2(2\tau)\theta_3(2\tau), \quad 
\theta_3^2=\theta_2^2(2\tau)+\theta_3^2(2\tau), \quad 
\theta_4^2=-\theta_2^2(2\tau)+\theta_3^2(2\tau), \nn \\
& & 2\theta_2^2(2\tau)=\theta_3^2-\theta_4^2, \qquad 
2\theta_3^2(2\tau)=\theta_3^2+\theta_4^2 .
\eea

The series representation of the Eisenstein series $E_2$, $E_4$ and $E_6$ are given by
\bea
E_2(q)&=&1-24\sum_{n=1}^{\infty}\frac{nq^n}{1-q^n} ,\\ \nn
E_4(q)&=&1+240\sum_{n=1}^{\infty}\frac{n^3 q^n}{1-q^n} ,\\ \nn
E_6(q)&=&1-504\sum_{n=1}^{\infty}\frac{n^5 q^n}{1-q^n} \, .\\ \nn
\eea
The functions  $E_4$ and $E_6$ can be written in terms of theta functions using the following 
expressions
\bea
E_4&=&\frac{1}{2}(\theta_3 ^8+\theta_4 ^8+\theta_2 ^8)  , \\ \nn
E_6&=&\frac{1}{2}(-\theta_2^6(\theta_3^4+\theta_4 ^4)\theta_2 ^2+\theta_3^6(\theta_4^4-\theta_2 ^4)\theta_3 ^2+\theta_4^6(\theta_3^4+\theta_2 ^4)\theta_4 ^2).  \\ \nn
\eea
Eisenstein series with the $U(1)$ chemical potential  are defined by 
\bea
E_{4,1}(z)&=&\frac{1}{2}(\theta_3 ^6 \theta_3 ^2(z)+\theta_4 ^6 \theta_4 ^2(z)+\theta_2 ^6 \theta_2 ^2(z))  , \\ \nn
E_{6,1}(z)&=&\frac{1}{2}(-\theta_2^6(\theta_3^4+\theta_4 ^4)\theta_2 ^2(z)+\theta_3^6(\theta_4^4-\theta_2 ^4)\theta_3 ^2(z)+\theta_4^6(\theta_3^4+\theta_2 ^4)\theta_4 ^2(z)) . \\ \nn
\eea
The decomposition of these series in terms of even and odd parts 
are defined by 
\bea
E_{4,1}=E_{4,1} ^{{\rm even}}\theta_{{\rm even}}+E_{4,1} ^{{\rm odd} }(z) \theta_{{\rm odd}}(z), \\ \nn
E_{6,1}=E_{6,1} ^{{\rm even} }\theta_{{\rm even} }+E_{6,1} ^{{\rm odd} }(z) \theta_{{\rm odd} }(z).
\eea
where
\bea
\theta_{{\rm even} }(z)=\theta_3(2\tau,2z) \qquad \theta_{{\rm odd} }(z)=\theta_2(2\tau,2z).
\eea
Any Jacobi form of index 1, $f_{s,1}(\tau,z)$ such as  $E_{4,1},\; E_{6,1}$) can be decomposed as:
\be
f_{s,1}(\tau,z)=f_{s,1} ^{{\rm even} }(\tau)\theta_{{\rm even}}(\tau,z)+
f_{s,1} ^{{\rm odd} }(\tau)\theta_{{\rm odd} }(\tau,z).
\ee
Then the definition of $\Gamma_{3,2} ^{(r,s)}\otimes f_{s,1}$ is iven by
\be
\Gamma_{3,2} ^{r,s}\otimes f_{s,1}=
\Gamma_{3,2} ^{r,s}({\rm even}) f_{s,1} ^{{\rm even} }+\Gamma_{3,2} ^{r,s}(odd) f_{s,1}^{{\rm odd}} \; ,
\ee
where
\bea
\Gamma_{3,2} ^{(r,s)} ({\rm even})
&=&\sum_{\begin{smallmatrix}m_1,m_2,n_2 \in \mathbb{Z},\\n_1=\mathbb{Z} + \frac{r}{N}, b \in 2\mathbb{Z}
\end{smallmatrix}}q^{\frac{p_L ^2}{2}}\bar{q}^{\frac{p_R ^2}{2}}e^{2\pi i m_1 s/N} \\ \nn
\Gamma_{3,2} ^{(r,s)}({\rm odd})
&=&\sum_{\begin{smallmatrix}m_1,m_2 ,n_2,  \in 
\mathbb{Z},\\ n_1= \mathbb{Z} + \frac{r}{N},  b \in 2\mathbb{Z}+1
\end{smallmatrix}}q^{\frac{p_L ^2}{2}}\bar{q}^{\frac{p_R ^2}{2}}e^{2\pi i m_1 s/N}.\\ \nn
\eea
where $p_L, p_R$ are given in (\ref{defomega}) and $N$ is the order of the $g'$ action. 

We now list the set of identities relating ${\cal E}_2$ and Eisenstein series as well 
as theta function which  have been used to obtain the results in this paper. 
First we have the identity 
\be
{\cal E}_2(\tau)^2=\frac{1}{4}(2\theta_3 ^8+2\theta_4 ^8-\theta_2 ^8), 
\ee
and we define ${\cal E}_2^2$ in  the presence of  the $U(1)$ chemical potential using 
the relation
\be
{\cal E}_{2,1}(\tau,z)^2=\frac{1}{4}(2\theta_3 ^6 \theta_3(z)^2+2\theta_4 ^6 \theta_4(z)^2-\theta_2 ^6 \theta_2(z)^2).
\ee
We have then the identity
\be
{\cal E}_{2,1}(\tau,z)^2(E_6+2{\cal E}_2(\tau) E_4)={\cal E}_{2}(\tau)^2(E_{6,1}+2{\cal E}_2(\tau) E_{4,1}).
\ee
These are the following identities between ${\cal E}_2$ and Eisenstein series 
at $2\tau$. 
\bea \label{rel}
E_6(2\tau)&=&\frac{1}{8}{\cal E}_2(\tau)(11 {\cal E}_2^2(\tau)-3E_4),\\ \nn
E_4(2\tau)&=&\frac{1}{4}(5 {\cal E}_2^2(\tau)-E_4).
\end{eqnarray}
We note that ${\cal E}_2^3$ can be rewritten in terms of Eisenstein series and 
a single power of ${\cal E}_2$ using the relation
\begin{equation}
{\cal E}_2^3(\tau)=\frac{1}{4}(E_6+3E_4 {\cal E}_2 (\tau)).
\end{equation}
Their modular transformed versions can be simplified as:
\bea \label{modrel}
E_6(\tau/2)&=&{\cal E}_2(\tau/2)(-11 {\cal E}_2^2(\tau/2)+12E_4),\\ \nn
E_4(\tau/2)&=&(5 {\cal E}_2^2(\tau/2)-4 E_4),\\ \nn
{\cal E}_2^3(\tau/2)&=&(-2E_6+3E_4 {\cal E}_2 (\tau/2)).
\eea
Finally we also quote the identities obtained in  
in \cite{Datta:2015hza} relating ${\cal E}_2$ and theta functions. 
\bea
-(\theta_3^8 \theta_4^4 + \theta_4^8 \theta_3^4 )&=& -\frac{2}{3} \left(E_6 + 2  {\cal E}_2(\tau) E_4 \right), \\ \nn 
\theta_3^8 \theta_2^4 + \theta_2^8 \theta_3^4  &=& -\frac{2}{3} \left(E_6 - {\cal E}_2(\frac{\tau}{2}) E_4 \right), \\ \nn
\theta_2^8 \theta_4^4 - \theta_2^8 \theta_4^4  &=& -\frac{2}{3} \left(E_6 -  {\cal E}_2(\frac{\tau+1}{2}) E_4 \right).
\eea

For simplifications in the section  \ref{su2bm} dealing with  the $2B$ orbifold we need
to relate theta functions and ${\cal E}_4$. This is given by
\begin{equation}
 \theta_4^4(2\tau) =-({\cal E}_2-2 {\cal E}_4).
\end{equation}
Finally we have the interesting identity relating the $(0,2)$ sector of the new supersymmetric
index for the $2B$ model given in (\ref{gprm02})
to Eisenstein series 
\begin{equation}\label{int02id}
 \Phi^{(0,2)}_{R^+}\theta_2 ^6+ \Phi^{(0,2)}_{NS^+}
 \theta_3^6- \Phi^{(0,2)}_{NS^-}\theta_4^6 =\frac{1}{3}E_6-\frac{4}{3}{\cal E}_2(\tau)E_4\; .
\end{equation}

\section{Threshold Integrals} \label{integrals}

In this appendix we detail the steps in performing the integral in 
(\ref{bbprimew}).  First we write the integrand in a from so that 
we can identity integrals which has already been performed. 
Adding and subtracting terms in the integrand we obtain
\begin{eqnarray}\label{thint0}
& & \Delta_G(T, U, V) - \Delta_{G'} ( T, U, V) = \int_{\cal{F}} \frac{d^2\tau}{\tau_2} 
\{ {\cal B}_G - {\cal B}_{G'} \} , \\ \nonumber
\qquad &=&  \int_{\cal{F}} \frac{d^2\tau}{\tau_2}  \left\{ 
\Gamma^{(0,0)} \otimes 2 (n_2-n_1)A(z) 
\right. \\ \nn
\qquad &&- \Gamma^{0,1} \otimes 
\left[24A(z)(\frac{n_1-12}{18})-12B(z){\cal E}_2(\tau)(\frac{2}{3}-\frac{\hat b}{2}) 
 \right] \\ \nn
\qquad &&-\Gamma^{(1,0)}\otimes \left[
24A(z)(\frac{n_1-12}{18})+6B(z){\cal E}_2(\frac{\tau}{2} )(\frac{2}{3}-\frac{\hat b}{2})\right]
\\ \nn
\qquad &&\left. -\Gamma^{(1,1)}\otimes \left[ 
24A(z)(\frac{n_1-12}{18})+6B(z){\cal E}_2(\frac{\tau+1}{2})(\frac{2}{3}-\frac{\hat b}{2})\right]
\right\}, \\ 
&=&- 24 \left(  (  \frac{1}{2} - \frac{3\hat b}{8} )  {\cal I}_1 + 
 ( \frac{n_1}{72} - \frac{1}{3} + \frac{\hat b}{8} ) ( {\cal I}_2 + {\cal I}_3)  \right),
\end{eqnarray}
where
\begin{eqnarray} \nonumber
{\cal I}_1 &=& \int_{\cal F}  \frac{\rm{d}^2\tau}{\tau_2} \left\{   \Gamma^{(0, 0)}_{3, 2} \otimes 
4 A(z) +  \Gamma^{(0, 1)}_{3, 2} \otimes \left( 
\frac{4}{3} A   - \frac{2}{3} B {\cal E}_2(\tau) \right) 
 + \Gamma^{(1,0)}_{3, 2} \otimes 
\left( \frac{4}{3} A + \frac{1}{3} B  {\cal E}_2(\frac{\tau}{2} ) \right)  \right. \\ \nonumber
& & \left.  + \Gamma^{(1,1)}_{3, 2} \otimes 
\left( \frac{4}{3} A +  \frac{1}{3} B  {\cal E}_2(\frac{\tau +1}{2} ) \right)  \right\} ,
\\ \nonumber
{\cal I}_2 &=& \int_{\cal F}  \frac{\rm{d}^2\tau}{\tau_2} \Gamma^{(0,0)}_{3, 2} 
\otimes  8A , \\ 
{\cal I}_3&=&\int \frac{\rm{d}^2\tau}{\tau_2}[\Gamma^{(0,0)}+\Gamma^{(0,1)}+\Gamma^{(1,0)}+\Gamma^{(1,1)}]\otimes  4A .
\end{eqnarray}
Using the results of the integrals in (\ref{intf1}) and (\ref{intf3}) in (\ref{thint0}) we obtain 
\begin{eqnarray}\label{intf4}
 \Delta_G(T, U, V) - \Delta_{G'} ( T, U, V) & =&  48
 \left(  (  \frac{1}{2} - \frac{3\hat b}{8} ) \log(\rm{det}({\rm{Im}}(\Omega))^6\left|\Phi_6(U,T,V) \right|^2  )   
 \right. \\ \nonumber
& &  +  ( \frac{n_1}{72} - \frac{1}{3} + \frac{\hat b}{8} ) 
 \log(\rm{det}
\left(\rm{Im}(\Omega))^{10}\left| \Phi_{10}(U,T,V) \right|^2 \right) \\ \nonumber
 && \left. +   ( \frac{n_1}{72} - \frac{1}{3} + \frac{\hat b}{8} ) \log(\rm{det}
\left(\rm{Im}(\Omega))^{10}\left| \Phi_{10}(2U,T/2,V) \right|^2 \right) 
 \right).
\end{eqnarray}

Let us first recall the results of one loop integration  or the theta lifts which are known 
from earlier work
\begin{eqnarray} \label{intf1}
{\cal I}_1  
&=&- 2 \log(\rm{det}
\left(\rm{Im}(\Omega))^{10}\left| \Phi_{10}(U,T,V) \right|^2 \right),
\\ \nonumber
{\cal I}_2 
&=&-2\log(\rm{det}(\rm{Im}(\Omega))^6\left|\Phi_6(U,T,V) \right|^2  ).
\end{eqnarray}
The first equation is the result for the theta lift of the elliptic genus of $K3$ and 
the second equation is the result for the theta lift of the elliptic genus of the 
$2A$ orbifold of  $K3$. 
The new integral which we need to obtain the difference of one loop 
gauge thresholds for the non-standard embeddings is the following 
\begin{eqnarray}\label{todoint}
{\cal I}_3=\int \frac{\rm{d}^2\tau}{\tau_2}[\Gamma^{(0,0)}+\Gamma^{(0,1)}+\Gamma^{(1,0)}+\Gamma^{(1,1)}]\otimes  4A .
\end{eqnarray}
To evaluate this integral we can use the general  result in \cite{justin3} for integrals of this 
form which we will now state.  Given the integral of the form 
\bea \label{int1}
\tilde{I}(U,T,V)&=&\sum_{r,s=0}^{N-1}\sum_{b=0}^1\tilde{I}_{r,s,b}\;,\\
\tilde{I}_{r,s,b}&=&\int_{\mathcal{F}}\frac{\rm{d}^2\tau}{\tau_2} 
\sum_{\begin{smallmatrix}m_1,m_2,n_2\in\mathcal{Z}\\n_1\in \mathcal{Z}+\frac{r}{N}\\ j\in 2\mathcal{Z}+b\end{smallmatrix}}q^{p_L^2/2}\bar{q}^{p_R^2/2}e^{2\pi ism_1/N}h_b ^{r,s},\\ \nn
h_b ^{r,s}(\tau)&=&\sum_{n\in Z-b^2/4}c_b ^{r,s}(4n)q^n,\\ \nn
F^{r,s}(\tau, z) &=&h_0 ^{r,s}(\tau)\theta_3(2\tau,2z)+h_1 ^{r,s}(\tau)\theta_2(2\tau,2z) \\ \nonumber
&=& \sum_{b=0,1}\sum_{n\in \mathbb{Z}/N, j\in 2\mathbb{Z} +b}c_b ^{r,s}(4n-j^2)q^n z^j\,,
\eea
with the condition 
\begin{equation}
c_{0}^{(r, s)} ( u) = 0  \quad {\rm for}\;  u<0, \qquad c_1^{(r, s)}( u) = 0 \quad {\rm for}\;  u<-1, 
\end{equation}
the result for the integral is given by 
\be
\tilde{I}(U,T,V)=-
2\log[\rm{det}\,Im\Omega^k]-2\log[\rm{det}\,\tilde{\Phi}(U,T,V)]-2\log[\rm{det}\,\bar{\tilde{\Phi}}(U,T,V)]\,,
\ee
where 
\bea \label{phitilde}
\tilde{\Phi}(U,T,V)&=&e^{2\pi i(\tilde{\al}U+\tilde{{\beta}}T+V)}\\ \nn
&&\prod_{b=0,1}\prod_{r=0}^{N-1}\prod_{\begin{smallmatrix}k'\in \mathcal{Z}+\frac{r}{N},l\in \mathbb{Z},\\ j\in 2\mathbb{Z}+b\\ k',l\geq0,\, j<0 k'=l=0\end{smallmatrix}}\left(1-e^{2\pi i(k'T+lU+jV)}\right)^{\sum_{s=0}^{N-1}e^{2\pi isl/N}c_b^{r,s}(4k'l-j^2)},
\eea
and 
\bea
\tilde{\beta}&=&\frac{1}{24N}Q_{0, 0},\\ \nn
\tilde{\al}&=&\frac{1}{24N}\chi(M)-\frac{1}{2N}\sum_{s=0}^{N-1}Q_{0,s}
\frac{e^{-2\pi is/N}}{(1-e^{2\pi is/N})^2},\\ \nn
Q_{r,s}&=& N(c_0^{r,s}(0)+2c_1 ^{r,s}(-1)), \\ \nn
Q_{0, 0}&=&\chi(M)=24.
\eea
Now examining the integral we have in (\ref{todoint}),  it can be seen that  we can use the 
above result to perform the integral.  Comparing the form in (\ref{int1}) and (\ref{todoint}) 
we see that we have $N=2$,  therefore $r, s \in \{0, 1\}$ and all the coefficients 
\begin{equation}
c^{r, s}_b ( u) = \frac{1}{2} c_b(u) .
\end{equation}
where $c_b(u)$  are the coefficients in the expansion of the elliptic genus of 
$K3$ which is given by 
\begin{equation}
8A(\tau, z) = \sum_{b=0,1}\sum_{n\in \mathbb{Z}, j\in 2\mathbb{Z} +b}c_b (4n-j^2)q^n z^j\,.
\end{equation}
Thus we have 
\begin{equation}
Q_{r, s} = 24,  \qquad \tilde \alpha = 2, \qquad \tilde\beta = \frac{1}{2}.
\end{equation}
We can further simplify the expression in (\ref{phitilde}) as follows
\bea \nn
\tilde{\Phi}(U,T,V)&=&e^{2\pi i(2U+T/2+V)}\prod_{b=0,1}\prod_{r=0}^{1}\prod_{\begin{smallmatrix}k'\in \mathbb{Z}+\frac{r}{2},l\in \mathbb{Z},\\ j\in 2\mathbb{Z}+b\\ k',l\geq0, j<0 k'=l=0\end{smallmatrix}}(1-e^{2\pi i(k'T+2lU+jV)})^{c_b^{r,s}(4k'l-j^2)}\\ \nn
&=&e^{2\pi i(2U+T/2+V)}\prod_{b=0,1} \left [ 
\prod_{\begin{smallmatrix}k'\in \mathbb{Z},l\in \mathbb{Z},\\ j\in 2\mathcal{Z}+b\\ k',l\geq0, j<0 k'=l=0\end{smallmatrix}}
(1-e^{2\pi i(2k'T/2+l(2U)+jV)})^{c_b(8k'l-j^2)} \right. \\ \nn
&&\qquad\qquad\quad \left. \times 
\prod_{\begin{smallmatrix}k'\in \mathbb{Z},l\in \mathbb{Z},\\ j\in 2\mathbb{Z}+b\\ k',l\geq0,\,j<0\\k'=l=0\end{smallmatrix}}(1-e^{2\pi i((2k'+1)T/2+l(2U)+jV)})^{c_b(4(2k' +1)l-j^2)}
\right],
\\ \nn
&=&e^{2\pi i(2U+T/2+V)}\prod_{b=0,1}\prod_{r=0}^{1}\prod_{\begin{smallmatrix}k'\in \mathbb{Z},&l\in \mathbb{Z}, &j\in 2\mathbb{Z}+b\\ k',l\geq0,& j<0& k'=l=0\end{smallmatrix}}(1-e^{2\pi i(k'T/2+l(2U)+jV)})^{c_b(4k'l-j^2)}\\ \nn
&=&\Phi_{10}(2U,T/2,V).\\
\eea
In the last line we have used the definition of $\Phi_{10}$ which is the theta lift 
of the elliptic genus of $K3$.  Thus the result of the integral in (\ref{todoint}) 
is given by 
\begin{eqnarray} \label{intf3}
{\cal I}_3&=&\int \frac{\rm{d}^2\tau}{\tau_2}[\Gamma^{(0,0)}+\Gamma^{(0,1)}+\Gamma^{(1,0)}+\Gamma^{(1,1)}]\otimes  4A , \\ \nonumber
&=& - 2 \log(\rm{det}
\left(\rm{Im}(\Omega))^{10}\left| \Phi_{10}(2U,T/2,V) \right|^2 \right).
\end{eqnarray}

\section{Mathematica files} \label{mathfiles}
There are 2 Mathematica files included in the supplementary attachments.
Both the Mathematica files begin with definitions of the 
 generalized theta functions, Dedekind eta function, Jacobi forms of index 1 and   Eisenstein series. 
\begin{enumerate}
\item {\bf z4wilson.nb}:  The partition function of the shifted $E_8\times E_8$ lattice together 
with the left moving bosonic partition function on $K3$ is written in terms of
generalized theta functions and compared with the the $(0,1)$ sector of
(\ref{abc}). 
\item {\bf relations.nb}: Different relations given in the appendix \ref{notation} and 
used in the main text are checked by $q$ expansions. The formula for $N_h-N_v$ as a function 
of $\hat b$ given in (\ref{bnhnv}) is checked against the general expression (\ref{nhnv}).
$N_h-N_v$ is also  evaluated for the $2B$ model.  
\end{enumerate}

\providecommand{\href}[2]{#2}\begingroup\raggedright\endgroup



\end{document}